%% file: PRD_Fermi_v1.tex
\documentclass[prd,10pt,twocolumn,showpacs,nofootinbib,superscriptaddress,amsmath,amssymb]{revtex4-1}

\usepackage{graphicx}
\usepackage{dcolumn}
\usepackage{bm}

\usepackage{setspace}
\usepackage[T1]{fontenc} 
\usepackage[varg]{txfonts}
\usepackage{soul}
\usepackage[]{natbib}
\usepackage[english]{babel}
\usepackage{amsmath}
\usepackage{ragged2e}
\usepackage{amssymb}
\usepackage{hyperref}
\usepackage{ccaption}
\usepackage{graphics,psfrag,rotating}
\usepackage{graphicx}
\usepackage{bm}
\usepackage{color}
\usepackage{float}
\usepackage{lineno}
\usepackage{caption}
\usepackage[dvipsnames,table]{xcolor}
\hypersetup{urlcolor=RedViolet,
	    citecolor=ForestGreen ,
	    linkcolor=RoyalBlue, 
	    colorlinks=true}
\definecolor{lightgray}{rgb}{0.9,0.9,0.9}	    
\definecolor{green}{rgb}{0,0.5,0}
\definecolor{red}{rgb}{1,0,0}
\definecolor{blue}{rgb}{0,0,0.5}

\newcommand{\lsim}   {\mathrel{\mathop{\kern 0pt \rlap
  {\raise.2ex\hbox{$<$}}}
  \lower.9ex\hbox{\kern-.190em $\sim$}}}
\newcommand{\gsim}   {\mathrel{\mathop{\kern 0pt \rlap
  {\raise.2ex\hbox{$>$}}}
  \lower.9ex\hbox{\kern-.190em $\sim$}}}

\newcommand{\mnras}{{MNRAS}}

\newcommand{\apjs}{{The~Astrophysical~Journal~Supp.~Series}}

\definecolor{myred}{RGB}{220, 20, 60}
\definecolor{myblue}{RGB}{60, 20, 220}
\definecolor{mygreen}{RGB}{60, 220, 20}

\begin{document}

\preprint{APS/123-QED}

\title{Dark Matter search in dwarf irregular galaxies with the \textit{Fermi} Large Area Telescope}

\author{V. Gammaldi}
\email{viviana.gammaldi@uam.es}
\affiliation{Departamento de F\'{i}sica Te\'{o}rica, Universidad Aut\'{o}noma de Madrid, Madrid, Spain $\&$  Instituto de F\'{i}sica Te\'{o}rica, UAM/CSIC, E-28049 Madrid, Spain}
 
 \author{J. P\'erez-Romero}
 \email{judit.perez@uam.es}
\affiliation{Departamento de F\'{i}sica Te\'{o}rica, Universidad Aut\'{o}noma de Madrid, Madrid, Spain $\&$  Instituto de F\'{i}sica Te\'{o}rica, UAM/CSIC, E-28049 Madrid, Spain}
 
 \author{ J. Coronado-Bl\'azquez}
 \email{javier.coronado@uam.es}
\affiliation{Departamento de F\'{i}sica Te\'{o}rica, Universidad Aut\'{o}noma de Madrid, Madrid, Spain $\&$  Instituto de F\'{i}sica Te\'{o}rica, UAM/CSIC, E-28049 Madrid, Spain}
 
 \author{M. Di Mauro}\email{dimauro.mattia@gmail.com}
 \affiliation{INFN, Istituto Nazionale di Fisica Nucleare, via P. Giuria, 1, 10125 Torino, Italy}

\author{E.~V.~Karukes}
\email{ekarukes@camk.edu.pl}
\affiliation{AstroCeNT, Nicolaus Copernicus Astronomical Center Polish Academy of Sciences,\\ul. Rektorska 4, 00-614 Warsaw, Poland}

\author{M.A. S\'anchez-Conde}
\email{miguel.sanchezconde@uam.es}
\affiliation{Departamento de F\'{i}sica Te\'{o}rica, Universidad Aut\'{o}noma de Madrid, Madrid, Spain $\&$  Instituto de F\'{i}sica Te\'{o}rica, UAM/CSIC, E-28049 Madrid, Spain}

\author{P. Salucci}
\email{salucci@sissa.it}
 \affiliation{SISSA, International School for Advanced Studies, \\
  Via Bonomea 265, 34136, Trieste, Italy}
 \affiliation{INFN, Istituto Nazionale di Fisica Nucleare - Sezione di Trieste, \\
 Via Valerio 2, 34127, Trieste, Italy}

\date{\today}
\smallskip
\begin{abstract}
We analyze 11 years of \textit{Fermi}-LAT data corresponding to the sky regions of 7 dwarf irregular (dIrr) galaxies. DIrrs are 
dark matter (DM) dominated systems, 
proposed as interesting targets for the indirect search of DM with gamma rays.  
The galaxies represent interesting cases with a strong disagreement between the density profiles (core vs. cusp) inferred from observations and numerical simulations.
In this work, we addressed the problem by considering 
 two different DM profiles, based on both the fit to the rotation curve (in this case a Burkert cored profile) and results from N-body cosmological simulations (i.e., NFW cuspy profile). We also include halo substructure in our analysis, which is expected to boost the DM signal a factor of ten in halos such as those of dIrrs. 
For each DM model and dIrr, we create a spatial template of the expected DM-induced gamma-ray signal to be used in the analysis of \textit{Fermi}-LAT data. 
No significant emission is detected from any of the targets in our sample. Thus, we compute upper limits on the DM annihilation cross-section versus mass parameter space. Among the 7 dIrrs, we find IC10 and NGC6822 to yield the most stringent individual constraints, independently of the adopted DM profile.  
We also produce combined DM limits for all objects in the sample, which turn out to be dominated by IC10 for all DM models and annihilation channels, i.e. $b\bar{b}$, $\tau^+\tau^-$ and $W^+W^-$. The strongest constraints are obtained for $b\bar{b}$ and are at the level of $\langle\sigma v \rangle \sim 7 \times 10^{-26}\text{cm}^{3}\text{s}^{-1}$ at $m_\chi\sim 6$ GeV. 
Though these limits are a factor of $\sim 3$ higher than the thermal relic cross section at low WIMP masses, they are independent from and complementary to those obtained by means of other targets. 

\begin{description}
\item[PACS numbers]
\end{description}
\end{abstract}

\pacs{Valid PACS appear here}
\maketitle


\section{Introduction}

Astrophysical and cosmological evidence suggests that non-baryonic cold dark matter (DM) constitutes $84\%$ of the matter density of the Universe \citep{Ade:2015xua, Aghanim:2018eyx}. Although the actual nature of DM is still unknown, weak interacting massive particles (WIMPs) are well-motivated DM candidates. They are predicted to annihilate or decay into Standard Model (SM) particles, whose decay and hadronization processes would produce secondary particles, such as cosmic rays, neutrinos and gamma rays \citep{Bertone:2005xv}. 
These messengers are expected to be observable 
in ground-based or satellite observatories, laying the groundwork for the indirect searches of DM. Either the expected DM signal could be disentangled from the background (here defined as the emission from any well-known astrophysical source) resulting in a DM hint, or the absence of an exotic signal provides constraints on the nature of the WIMP particle, here its mass and annihilation cross section. Among the messengers, gamma rays have represented the golden channel as of today: they are (very-) high-energy neutral particles traveling practically undeflected, along straight paths in the Universe. 
Nonetheless, only an agreement of a few hints in several observation channels - i.e. the multi-messenger approach - would result in a competitive claim in the sense of the indirect detection of DM \cite{Bergstrom2013, MiniReview}.\\
\\
DM-dominated systems - e.g. galaxy clusters, dwarf spheroidal (dSph) galaxies as well as the Galactic center - are benchmark targets for indirect searches of DM ( see e.g. \cite{Conrad:2017pms, Charles:2016pgz} and refs therein). 
Among other targets, Milky Way dSph galaxies are considered to be especially promising objects due to their relatively close position and their appearance as point-like or marginally extended sources in gamma-ray telescopes.  
Although the DM density profiles inferred from their kinematics are often affected by large uncertainties \citep{triaxSph}, the contamination from intrinsic gamma-ray sources is negligible in this type of objects, making them particularly appealing as well \cite{Winter:2016wmy}. \\ 
\\
Recently, dwarf irregular (dIrr) galaxies in the Local Volume have been claimed to be 
interesting targets for gamma-ray DM searches as well: (i) unlike pressure-supported dSph galaxies, dIrrs are rotationally-supported galaxies, i.e. their DM profiles can be obtained from their rotation curves (RC) and, indeed, from these they appear to be DM-dominated systems at all radii \citep{Oh:2015xoa,oh1,gentile06} - just as dSphs;  
(ii) dIrrs are isolated galaxies of the Local Volume with DM halo mass $M_{200}\approx 10^7-10^{10}\,M_\odot$. Here, we consider dIrrs at a distance less than $\sim 1$ Mpc. Roughly speaking, the astrophysical component of the gamma-ray DM flux scales with $\approx M_{200}^2/d^2$, so that it is expected to have comparable contributions for both the dIrrs and dSphs (which typical have $M_{200}\approx 10^5-10^{7}\, M_\odot$ and $d<0.5$ Mpc), as observed in our analysis 
 (see e.g. Fig. 5 in \cite{Gammaldi:2017mio}); (iii) Unlike aged, largely evolved dSph galaxies, dIrrs are star-forming galaxies; yet the astrophysical gamma-ray emission from their star-forming regions has been estimated to be several orders of magnitude lower than the gamma-ray flux expected from DM annihilation 
events in the halo, for vanilla WIMP models. In other words, the astrophysical background emission is expected to be negligible in dIrrs similarly to the case of dSphs \cite{Gammaldi:2017mio}. In fact, dIrrs have star-forming regions of very small angular size (unresolved by most of gamma-ray observatories, e.g. Fig. 6 in \cite{Gammaldi:2017mio}), while their DM halos appear to be extended - due to their distances and typical values of the virial radius \cite{Gammaldi:2017mio}. \\
\\
Despite these advantages, dIrrs have not been studied so far with data of the Large Area Telescope on board the NASA \textit{Fermi} satellite (\textit{Fermi}-LAT) \cite{atwood2013pass} in the context of DM searches. The \textit{Fermi}-LAT is a pair conversion telescope designed to observe the energy band from 20 MeV to greater than 300 GeV, that has been surveying the sky searching for gamma-ray sources since 2008~\cite{2009ApJ...697.1071A}. Several point-source \textit{Fermi}-LAT catalogs have been released that contain hundreds to thousands of gamma-ray objects, many of them previously unknown \cite{Fermi-LAT:2019yla}. In this work, we will perform the first spatial analysis of 7 dIrr galaxies in the Local Volume with \textit{Fermi}-LAT data. \footnote{
A recent work \cite{Hashimoto:2019obg} focuses on Low Surface Brightness galaxies (LSBs) to constrain the DM model with \textit{Fermi}-LAT data. Such objects are similar to dIrrs but differ in key aspects, particularly in morphology: both LSB and dIrr galaxies are rotationally supported systems, but LSBs still show spiral arms, while dIrrs are much more irregular.  Further, 
the LSB sample in Ref.~\cite{Hashimoto:2019obg} includes (i) farther objects, with distances $> 20$ Mpc and already significant redshifts, while dIrrs are Local Volume objects with distances $ \lesssim 1$ Mpc; (ii) J-factors of $10^{14}-10^{16}\,\text{GeV}^{2}\text{cm}^{-5}$, i.e. a factor 100 - or more - smaller than the J-factors in this work (for details see section \ref{modeling}). Secondly, the LSB in Ref.~\cite{Hashimoto:2019obg} are point-like sources for \textit{Fermi}-LAT, while dIrrs in this work are extended sources due to their proximity. For the same reason, there is no overlapping between the list of objects in the LSBs analyses (e.g. \cite{Hashimoto:2019obg, 2021MNRAS.501.4238B}) and our work. Instead, an overlap in the targets exists with some preliminary analyses obtained with the High Altitude Water Cherenkov (HAWC) observatory \cite{Cadena:2017ldx, Cadena:2019lor}, as well as with the very recent study of WLM published by the High Energy Stereoscopic System \cite{2021arXiv210504325H}, both of them at TeV energy scale. Indeed, the results of these studies are complementary to this work.}. 
We will also discuss in-depth both the main theoretical advantages and uncertainties related to this new class of sources, as briefly introduced in points (i)-(iii) above. In more details:  \\
\\
(i) Due to the current uncertainty that affects the inner DM density profile (in particular for galaxies in the range of masses considered in this paper), we adopt - for each dIrr galaxy in our sample - both a core and a cuspy profile, 
as we discuss in Section \ref{Irr}.\\
 \\
(ii) For the computation of the expected gamma-ray flux, we include in our analysis 
the effect of halo substructure, leading to the so-called "subhalo boost". In fact, 
subhalos may play a crucial role for indirect DM searches, in terms of an enhancement in the gamma-ray flux. In particular, the boost in the gamma-ray flux due to substructures is expected to be more dominant in field halos (i.e dIrrs, that are main halos of the Local Volume), instead of dwarf satellites (e.g. dSphs, that fill subhalos in our galaxy) \cite{Ando:2019xlm}, as we discuss in Section \ref{modeling}.
\\
\\
(iii) Due to both the apparent angular size of the dIrr DM halos as well as the inclusion of halo substructure - that particularly boosts the annihilation flux in the outer regions of host halos 
- the DM signal from dIrrs could appear as extended for \textit{Fermi}-LAT. For this reason, we perform a spatial analysis of these objects, instead of the point-like analysis commonly adopted in the study of dSphs, as 
discussed in section \ref{modeling}. 
\\
\\
In section \ref{gammarayDMsection} 
we present the \textit{Fermi}-LAT data analysis for each individual target in our sample, by using 11 years of LAT data and adopting the corresponding spatial template for each of the dIrrs. 
No significant emission is detected from any of the targets. Indeed, the absence of a gamma-ray signal detection in our sample of dIrrs represents one important result of this work, 
which - in addition - is in agreement 
with the expectation \cite{Gammaldi:2017mio}. 
Thus, we proceed and compute $95\%$ confidence level (C.L.) upper limits in the standard annihilation cross section versus WIMP mass parameter space, $<\sigma v> - m_\chi$, for each dIrr and considered DM model scenario. 
The results of both the individual and combined analysis of our sample of dIrrs are presented and discussed in Section \ref{sec:DMlimits}. 
Finally, we summarize and discuss the main results of this work in Sec. \ref{conclusions}. 

\section{Kinematics and DM modelling}
\label{Irr}

DIrrs are rotationally-supported galaxies and DM-dominated systems at all radii, with masses $M_{200}\approx 10^7-10^{10}\,M_\odot$. Yet, the exact shape of the inner DM density profile in dIrr galaxies still represents an open issue. 
On the one hand, the cold HI and $H_\alpha$ rotating disks in dIrrs yield high resolution and high quality RCs, which point 
to the existence of cores in their centers, with halos extending well beyond their star-forming regions \citep{urc1}. On the other hand, this kind of cored-like profiles are in contrast with the predictions of  
 N-body, DM-only cosmological simulations, which predict cuspy DM profiles such as the Navarro-Frenk-White (NFW) \cite{1996ApJ...462..563N, 1997ApJ...490..493N}.  
Nevertheless, the current DM profiles around galaxies could be different from the primordial ones and/or from the results of DM-only simulations, due to e.g. baryonic feedback \cite{Gomez-Vargas:2013bea, Schaller:2014uwa, 2017MNRAS.472.2153P}.
Unfortunately, the impact of baryons in cosmological simulations at the scales relevant for this work is not yet fully understood.
Whether or not a DM core will form in a given dwarf galaxy may depend primarily on its stellar mass to halo mass ratio, e.g.~\cite{Read:2018fxs, DiCintio:2014xia}. 
In fact, bursty star formation limited to dense $H_2$-rich regions creates repeated,
fast outflows which break the adiabatic approximation. These fast and repeated outflows progressively
lower the central DM density of galaxy halos and turn DM
central cuspy profiles into much shallower cores \cite{Read:2018fxs, DiCintio:2014xia, Sawala:2015cdf, 2012MNRAS.422.1231G}. Furthermore, the stars can be dynamically heated
similarly to the DM, leading to a stellar velocity dispersion
that approaches the local rotational velocity of the stars ($V/\sigma \sim 1$) within the projected half radius of the stars. Nonetheless, some authors \cite{Bose:2018oaj, 2019MNRAS.488.2387B} found that the fast and repeated outflows are not able to 
lower the central DM density of galaxy halos, and the cuspy profiles remain unperturbed. This is 
probably due to a relatively low gas density threshold for converting gas into stars which prevents the gas from becoming gravitationally dominant on kpc scales.\\
In light of this debate, we prefer to include in our analysis both a core and cuspy profile, in this way also estimating how such an open issue may affect our final results. First, we study the RCs of the sample of seven dIrr galaxies that will be later analyzed in gamma rays. Four of them (NGC6822, IC10, Wolf-Lundmark Melotte (WLM) and IC1613) have been selected as the most promising objects \footnote{The four most promising objects have been selected based on the results of \cite{Gammaldi:2017mio} - i.e. the first proof-of-concepts paper. In the latter, a point-like study based on a theoretical approach was developed, by taking into account several key factors, e.g. not only the mass, distance and angular dimension of the targets, but also their position in the sky and their star-forming gamma-ray emission, expected to be negligible with respect to both the gamma-ray emissivity expected by the DM halos and the diffuse and isotropic components of the background.} of thirty-six dIrr galaxies, that were previously investigated as a new category of targets for indirect DM searches \cite{Gammaldi:2017mio}; the last three galaxies (Phoenix, DDO210 and DDO216) are brand new targets, never studied before in this context and considered as interesting targets due to their mass and distance. The seven dIrrs are located at a distance less than $\sim 1$ Mpc, and are thus in the Local Group. Starting from the distance of $\sim 4$ Mpc, dIrrs start to outnumber dSphs, indeed the total number of known dIrrs in the Local Volume is at present $\sim 200\%$ of that of known dSphs \cite{Gammaldi:2017mio}. In this regard, we note that our sample only includes dIrrs located in a particular region of the sky, due to availability of data \cite{2017MNRAS.472.3761N,Karachentsev:2013cva,Hunter:2004nw,Hidalgo:2013zxa,Hunter_2012}.  
 The main characteristics of these galaxies are shown in Table \ref{tab:properties}, which provides name, distance, scale length $R_D$, stellar mass $M_D$ and position in the sky. Here the stellar disk scale length is defined as the radius at which the surface luminosity 
of a galaxy has fallen off by a factor of $e$ ($\sim$2.7).\\

First of all, we fit the observed RC data (see Appendix~\ref{AppA}) to a parametrized model described below and perform the global mass modelling. We consider the contribution to the total measured circular velocity from two components: the luminous part (stellar disk and the gaseous component when available) and the DM contribution, i.e.: 
\begin{table}[H]
\centering
\begin{tabular}{ | c | c | c | c | c | c | c |}
\hline
 Name & Distance & $R_{D}$ & $\log_{10}M_{D}$ & l & b & RC \& $M_{D}$ \\
      &   [Mpc] & [kpc]  &  [$\rm M_{\odot}$]  & [deg] & [deg] & reference   \\
\hline
\rowcolor{lightgray} 
NGC6822 & 0.48 & 0.66 & 7.0 & 23.3 & -18.4 & \cite{2017MNRAS.472.3761N}\\
IC10    & 0.79& 0.79  & 8.1 & 119.0 & -3.3 & \cite{Oh:2015xoa}  \\ 
\rowcolor{lightgray}
WLM     & 0.97  & 0.55 & 7.2 & 75.9 & -73.6 &\cite{Oh:2015xoa}  \\
IC1613  & 0.76  & 0.64 & 7.5 & 129.7 & -60.6 &\cite{Oh:2015xoa}  \\
\rowcolor{lightgray}
Phoenix & 0.44 & 0.23 & 6.8 & 272.2 & -68.9 &\cite{Kacharov_2016, Shao_2018}\\ 
DDO210  & 0.9          & 0.17 & 5.8 & 34.0 & -31.3 & \cite{Oh:2015xoa} \\
\rowcolor{lightgray}
DDO216  & 1.1           & 0.54& 7.2 & 61.5 & -67.1 & \cite{Oh:2015xoa}\\
\hline
\end{tabular}
\caption{\centering \footnotesize{Sample of the seven dIrr galaxies studied in this work, together with their distances ($D_{\rm Earth}$), scale lengths, $R_D$, and stellar masses, $M_D$,  and position in the sky. In the last column we provide for each target the reference for both the Rotation Curve (RC) data (see also Appendix~\ref{AppA}) and stellar mass $M_D$ data \cite{2017MNRAS.472.3761N,Karachentsev:2013cva,Hunter:2004nw,Hidalgo:2013zxa,Hunter_2012}.}}
\label{tab:properties}
\end{table}
\begin{equation}
    v^2_{tot}(r) = v^2_{disk}(r)+v^2_{gas}(r)+v^2_{dm}(r). 
\label{eqv:vel tot}
\end{equation}

We assume that the stellar component, $v_{disk}$, is well represented by a Freeman disk \cite{1970ApJ...160..811F} and, thus, can be written as:

\begin{equation}
v^2_{disk}(r)=\frac{1}{2}\frac{G M_{D}}{R_{D}} \left(\frac{r}{R_{D}}\right)^2 (I_{0} K_{0}-I_{1} K_{1}), 
\label{eqn:freemandisk}
\end{equation}

\noindent
where $M_{D}$ is the stellar mass, $R_D$ is the disk scale length, and $I_{n}$ and $K_{n}$ are the modified Bessel functions computed at $\frac{r}{2\,R_{D}}$. The disk scale length and the distribution of the gas component are taken from literature and listed in Table~\ref{tab:properties}.\footnote{This is done for all dIrrs but Phoenix~I, for which data for the gas distribution are not available. Thus, for this object we assume that the gas follows the Freeman disk given by Eq.~(\ref{eqn:freemandisk}), yet with $R_D$ being three times that of the disk~\cite{Tonini:2005je}. We take the gaseous mass to be $\log_{10}M_{\rm HI}=4.92$ from \cite{Shao_2018}.}
 
 The stellar mass is left as a free parameter: in Table~\ref{tab:properties} we show the values available in literature that we adopt as central values for the Gaussian priors with the standard deviation equal to 0.1\footnote{If we adopt larger standard deviation the implied stellar masses become unrealistically low/high.}.

In more detail, our model of Eq.~(\ref{eqv:vel tot}) has three free parameters: $M_D$ and two parameters (the core density and scale radius) describing the DM density profile, either Burkert ($\rho_c , \, r_c$) or NFW ($\rho_0 ,\, r_s $) respectively. We explore the model parameter space using uniform priors for the DM parameters and Gaussian priors for $M_D$\footnote{The good quality of the RC of NGC6822, allows to leave the stellar mass free.}. The posterior is then sampled using \texttt{emcee}, an open source affine-invariant Markov chain Monte Carlo (MCMC) ensemble sampler~\cite{2013ascl.soft03002F}. 

The best-fit maximum likelihood (ML) values of the MCMC analysis for the Burkert profile are presented in  
Table~\ref{tab:parameters_bur}, namely the core density, $\rho_c$, and the core radius, $r_c$: 
 
\begin{table*}[ht!]
\centering
\begin{tabular}{ | c | c | c | c | c | c | c | c |}
\hline
Name & $r_c$ & $\log_{10}\rho_c $ & $\log_{10}M_D$ & $\chi^2_{\mathrm{red}}$ & $ R_{200}$ &  $\log_{10}M_{200}$ & $\theta_{200}$\\ 
& $\text{[kpc]}$  & [$\rm M_{\odot}/kpc^3$] & [$\rm M_{\odot}$] & --- & [kpc] & [$\rm M_{\odot}$] & [deg] \\ 
\hline
\rowcolor{lightgray} 
NGC6822 & $3.3^{+0.8}_{-0.7}$ & $7.5\pm0.1$ & $7.9^{+0.2}_{-0.3}$ & $0.1$ & $62.9^{+8.4}_{-6.7}$  &$10.5^{+0.2}_{-0.1}$ & $7.5$\\ 
IC10 & $2.0^{+0}_{-1.5}$ & $8.2^{+0.4}_{-0.2}$ & $^*8.1\pm0.1$ & $0.1$ & $71.3^{+7.1}_{-47.6}$ & $10.6^{+0.1}_{-1.5}$ & $5.2$\\ 
\rowcolor{lightgray}
WLM &$1.3^{+0.2}_{-0.1}$ & $7.8\pm0.1$ & $^*7.1^{+0.5}_{-0.9}$ & $0.1$ &$33.3^{+2.0}_{-1.5}$ &$9.6\pm0.1$ & $2.0$ \\ 
IC1613 & $7.0^{+0}_{-1.2}$  & $6.3\pm0.05$ & $^*7.1^{+0.07}_{-0.06}$ & $0.9$  & $45.7^{+1.7}_{-6.9}$   &  $10.0^{+0.05}_{-0.2}$ & $3.4$\\ 
\rowcolor{lightgray}
Phoenix & $0.2$  & $7.5$ & $^*6.8$ & $1.2$  & $3.6$   &  $6.7$ & $0.5$\\ 
DDO210 & $0.5^{+0.9}_{-0.2}$  & $8.0^{+0.1}_{-0.2}$ & $^*5.8\pm0.1$ & $0.5$  & $14.2^{+23.7}_{-4.6}$   &  $8.5^{+1.2}_{-0.5}$ & $0.9$\\ 
\rowcolor{lightgray}
DDO216 & $0.3^{+0.3}_{-0.1}$  & $8.1\pm0.3$ & $^*7.2\pm0.1$ & $0.3$  & $10.8^{+3.7}_{-2.7}$   &  $8.2\pm0.3$ & $0.6$ \\  
\hline
\end{tabular}
\caption{\centering \footnotesize{Best-fit ML values and corresponding 1~$\sigma$ uncertainties for the main parameters of our sample of dIrrs, for a Burkert DM density profile. Asterisks indicate that the estimation of the stellar mass was taken from \cite{Oh:2015xoa} and the Gaussian prior is further assumed. We provide both the core radius $r_c$ and density $\rho_c$, the stellar mass $M_D$, the $\chi^2_{\mathrm{red}}$ value, $R_{200}$ and $M_{200}$. Finally, $\theta_{200}$ is the angular extension of the galaxy, defined as the sky angle subtended by $R_{200}$. Note that we do not provide the uncertainties for the Phoenix galaxy due to very poor quality of its RC data.}}
\label{tab:parameters_bur}
\end{table*}

\begin{equation}
    \rho_\mathrm{Bur}(r) = \frac{\rho_c\,r_c^3}{(r+r_c)\,(r^2+r_c^2)}.
\label{eqn:Burkert}    
\end{equation}
In Table~\ref{tab:parameters_bur} we also list values of the reduced $\chi^2_{\mathrm{red}}$, where the number of d.o.f. is the number of RC data points minus the number of free parameters (in the Burkert profile case is equal to 3). We fit the RC of each galaxy, instead of adopting the Universal Rotation Curve hypothesis as in \cite{Gammaldi:2017mio, 2017MNRAS.465.4703K}. This fact generates small differences with respect to \cite{Gammaldi:2017mio}, in the resulting parameters of the Burkert profiles and the related uncertainties.

We follow the same procedure for the NFW profile:

\begin{equation}\label{eqn:NFW}
 \rho_\mathrm{NFW}(r)=\frac{\rho_0}{\left(\frac{r}{r_s}\right)\left(1+\frac{r}{r_s}\right)^{2}},
\end{equation}
where $\rho_0$ and $r_s$ are the density normalization and the scale radius, respectively. However, even in those cases where this profile shows reasonable fits to the RC data, the obtained best-fit values for the concentration parameter are unrealistically low for a $\Lambda$CDM Universe~\cite{McGaugh_2003,Sanchez-Conde:2013yxa,Read_2016}. We also tried to use uniform priors based on structure formation models \cite{Ando:2020yyk}, with similar results. 
Consequently, either the DM in these galaxies is distributed somewhat differently from what is expected from the DM-only $\Lambda$CDM cosmological simulations, or the data contains unknown systematic uncertainties that should be taken into account, e.g. systematic uncertainties related to the inclination and/or distance errors in the RC reconstruction \cite{10.1093/mnras/stw1251}. 
Thus -- by looking out for an agreement with $\Lambda$CDM cosmology and the DM-only simulations -- we still model each dIrr galaxy with an NFW profile but with one key assumption: $M_{200}$, the mass contained within the virial radius, $R_{200}$, and obtained by the fit of the RCs for the Burkert profiles (Table \ref{tab:parameters_bur}), is also taken as a starting point to build the NFW profiles. Though valid, just as a first approximation, this assumption is good enough for our purposes, especially when considering that, later in our work, we will define and use a set of four different DM models (described in the next section) to conservatively encapsulate the uncertainties coming from the modelling of the DM distribution in our sample of dIrrs. Thus, uncertainties from the assumption of $M^{\rm NFW}_{200}=M^{\rm Bur}_{200}$ are subdominant compared to the level of global uncertainties from the envelope of all four considered DM models \cite{urc1}.
From this value of $M^{\rm NFW}_{200}$ for each dIrr, it is possible to obtain the corresponding radius $R_{200}$:
\begin{equation}
R_{200}=\left(\frac{3 M_{200}}{4\pi \Delta_{200}\rho_{crit}}\right)^{1/3},
\label{eq:R200}
\end{equation}
$\Delta_{200}$ being the overdensity with respect to the critical density of the Universe, $\rho_{crit}\,=\,137\,$M$_{\odot}\,$kpc$^{-3}$. 

The next step to build the NFW profile is to obtain the halo concentration. For this purpose we use the parametrization for main halos \cite{Sanchez-Conde:2013yxa}:
\begin{equation}
    c_{200}(M_{200},z=0)=\sum_{i=0}^5 c_i \times \left[\ln\left(\frac{M_{200}}{h^{-1}M_\odot}\right)\right]^i,
    \label{eq:c_M}
\end{equation}
which has proven to work especially well for objects in the mass range between dSphs and galaxy clusters, and includes the flattening of $c(M)$ at lower halo masses, first pointed out by these authors and now widely accepted \cite{Ludlow:2016ifl, Wang:2019ftp}. Together with the value of $R_{200}$ given by Eq.~\ref{eq:R200}, it is now easy to obtain the scale radius, $r_s$:
\begin{equation}\label{eqn:scale radius}
r_s \equiv R_{200}/c_{200}.
\end{equation}
Finally, we impose the condition
\begin{equation}
M_{200}=\int_0^{R_{200}}\rho_{\rm NFW}(r)r^2drd\Omega
\end{equation}
and we get 
$$ \rho_0 = \frac{2~\Delta_{200}~\rho_{crit}~c_{200}}{3~f(c_{200})},$$
where $f(c_{200})=\frac{2}{c_{200}^2}\left(\ln{(1+c_{200})}-\frac{c_{200}}{1+c_{200}}\right)$.
The variable that we will use to define the spatial extension of the dIrrs is $\theta_{200}$, i.e. the sky angle subtended by $R_{200}$:
\begin{equation}\label{eqn:theta}
    \theta_{200} = \arctan\left(\frac{R_{200}}{D_{\rm Earth}}\right).
\end{equation}
The resulting NFW profile parameters are given in Table \ref{tab:parameters_nfw}. The RCs corresponding to both the data-consistent Burkert profile and the $\Lambda $CDM-consistent NFW profile are left for Appendix~\ref{AppA}. Despite the adopted theoretical model, the NFW profile remains a consistently worse than the Burkert profile, for reasons we gave above, i.e. the main discrepancies between the model and the data for some objects may be associated with (i) an incomplete knowledge of baryonic effects in numerical simulations and/or (ii) systematic uncertainty in the determination of the RC data, e.g., the estimated inclination angle of the galaxy. Due to these unknowns, the quality of the RC fit was not considered as primary when choosing the list of Irrs from \cite{Gammaldi:2017mio} for this analysis. Instead, we preferred the objects with the most promising J-factors (see \cite{Gammaldi:2017mio} and Section \ref{modeling} ). Nonetheless, 5 of 7 RCs are well fitted with a cored profile, by including the baryon component.  

\begin{table}[H]
\centering
\begin{tabular}{ | c | c | c | c | c | c |}
\hline
 Name & $c_{200}$ & $r_s$ & $\log_{10}\rho_0$ & $R_{200}$ & $\theta_{200}$ \\
 &  & [kpc] & [$\rm M_{\odot}/kpc^3$] & [kpc] & [deg]\\
\hline
\rowcolor{lightgray} 
NGC6822 & 10.7 & 5.9 & 6.9 & 62.6 & 7.4\\
IC10 & 10.4 & 6.8 & 6.8 & 70.3 & 5.1\\ 
\rowcolor{lightgray}
WLM & 12.2 & 2.8 & 7.0 & 33.6 & 2.0\\
IC1613 & 11.4  & 4.0 & 6.9 & 45.7 & 3.4\\ 
\rowcolor{lightgray}
Phoenix & 18.7  & 0.2 & 6.9 & 3.5 & 0.5\\
DDO210 & 14.5  & 1.0 & 7.2 & 14.5 & 0.9\\
\rowcolor{lightgray}
DDO216 & 15.3  & 0.7 & 7.3 & 10.8 & 0.6\\  
\hline
\end{tabular}
\caption{\centering \footnotesize{NFW density profile parameters for our sample of dIrrs, obtained assuming $M_{200}^{\rm Bur}=M_{200}^{\rm NFW}$; see the text for details, and Eqs.~(\ref{eqn:NFW}--\ref{eqn:theta}) for the definition of each parameter. The concentration $c_{200}$ corresponds to the one provided by the concentration-mass ($c-M$) relation in \cite{Sanchez-Conde:2013yxa} for $\mathrm{\Lambda CDM}$ halos.}}
\label{tab:parameters_nfw}
\end{table}

\section{Annihilation signal and inclusion of halo substructure}
\label{modeling}

The DM modelling performed in the previous section is used as the starting point to obtain the induced DM annihilation gamma-ray flux from these objects. Assuming that the DM is composed of WIMPs \cite{Bertone:2004pz, Hooper:2009zm}, we can compute this expected flux as
\begin{equation}\label{eq:dm-flux}
\frac{d\phi_{\gamma}}{dE}(E, \Delta\Omega, l.o.s)=\frac{1}{4\pi}\frac{\langle\sigma v\rangle}{\delta\, m^2_{\chi}}\frac{dN_{\gamma}}{dE}(E)\times J(\Delta\Omega, l.o.s),
\end{equation}
where: $\frac{d\phi_{\gamma}}{dE}$ is the DM annihilation gamma-ray flux; $\langle\sigma v\rangle$ is the thermally-averaged annihilation cross-section; $\delta = 2$ if we assume Majorana DM particles and $\delta = 4$ if Dirac particles; $m_{\chi}$ is the DM mass; $\frac{dN_{\gamma}}{dE}$ is the WIMP annihilation spectrum; $J(\Delta\Omega, l.o.s)$ is the so-called astrophysical J-factor (computed along the line of sight (l.o.s.), and within a given solid angle $\Delta\Omega$). It is worth to notice that we can identify two main dependencies in the flux: first, the energy dependence, that appears only in the so-called particle physics term (which contains all the information about the mass of the DM candidate and the possible annihilation channels); secondly, the spatial dependence, appearing only in the J-factor. This allows to factorize these two terms independently, and implies that the spatial distribution of the DM is independent of the energy. The J-factor is then defined as 
\begin{equation}\label{eq:j-factor}
J(\Delta\Omega, l.o.s) = \int_0^{\Delta\Omega}d\Omega\int_{l.o.s}\rho^2(r)dl,
\end{equation}
where $\Delta\Omega=2\pi(1-\cos\alpha_{int})$, being $\alpha_{int}$ the integration angle and $\rho(r)$ the DM density profile.\\
From N-body cosmological simulations - which follow the structure formation processes according to $\Lambda$CDM (e.g.,~\cite{Kuhlen:2012ft, Zavala:2019gpq} and references therein) - , we expect the smallest structures or halos to form first. Then, via accretion or collapse, the bigger structures are formed, leaving a population of small structures - known as subhalos - within the main halos. Given the typical masses and sizes of dIrrs, we expect them to host a significant number of subhalos. The population of subhalos in a main halo can be parametrized as
\begin{equation}\label{eq:subhalos-distribution}
\frac{d^3N}{dVdMdc}= N_{tot}\frac{dP_V}{dV}(R)\frac{dP_M}{dM}(M)\frac{dP_c}{dc}(M,c),
\end{equation}
where: $N_{tot}$ is the total number of subhalos; $P_i$ with $i = V, M, c$ is the probability distribution in each of the domains normalized to 1; $V$ referring to the subhalo distribution in the volume of the main halo \footnote{However, the only real dependence is on the relative distance of the subhalos to the center of the host, so in the following we will refer to this distribution in volume as the Subhalo Radial Distribution.}; 
 $M$ to the subhalo mass and $c$ to subhalo concentration. This allows to model the subhalo distribution for each domain independently from one another (spatial distribution, mass and concentrations). The effect of taking into account all these small structures is to enhance the expected DM flux, which is usually quantified in terms of the so-called substructure boost factor, $B$. This boost can range from $B=0$, where the contribution of the substructure is absent, 
 up to $\sim 2$ orders of magnitude in the expected annihilation DM signal, depending on the details of the subhalo population and of the host halo mass (e.g.~\cite{Moline:2016pbm, Ando:2020yyk}). Given the impact that the substructure can have in the J-factors and its still uncertain nature (e.g., minimum mass to form clumps \cite{Green:2005fa, Bringmann:2009vf, Cornell:2013rza}, tidal stripping, subhalo survival or precise shape of subhalo DM density profiles \cite{Zavala:2019gpq}), it becomes convenient to cover the range of different but possible scenarios. Hereafter we proceed to describe the models that we use in this work to model the substructure within our sample of dIrrs:
\begin{itemize}
    \item $\frac{dP_V}{dV}$: Known as the Subhalo Radial Distribution (SRD), there is some variety of SRD models available in the community. The most used are the ones described in \cite{Springel:2008cc} and \cite{Diemand:2008in}, both based to N-body simulation results. Yet, as seen in section \ref{Irr}, dIrr galaxies agree better with data-driven DM profiles, like the Burkert profile, rather than with simulation-motivated profiles like the NFW. Because of this, in this work we decide to adopt a hybrid approach, for which, as in \cite{Springel:2008cc}, we choose the SRD to follow the profile of the main halo, i.e. Burkert or NFW depending on the considered model.
    
    \item $\frac{dP_M}{dM}$: The distribution in mass of the population of subhalos is known as the Sub-Halo Mass Function (SHMF). Once again, the main input for these models are the N-body simulations, which usually follows
    \begin{equation}\label{eq:subhalo-mass-function}
    \frac{dN}{dM}\propto M^{-\alpha}.
    \end{equation}
    In the literature the most common values are $\alpha = 1.9$ \cite{Springel:2008cc} and $\alpha = 2.0$ \cite{Diemand:2008in}. As shown in \cite{Moline:2016pbm}, the slope plays a crucial role on the boost factor calculation, since higher boosts are obtained for higher $\alpha$ values, i.e. for the case of a more numerous population of small subhalos. In order to cover all possible physical scenarios, we consider both cases,
    noting that the $\alpha = 1.9$ probably yields the more realistic J-factors as it is more in line with the latest results in the N-body simulation side \cite{Zavala:2019gpq}, while $\alpha = 2.0$ will provide an upper bound to the substructure boost values.
    \item $\frac{dP_c}{dc}$: 
    Even though in the literature it is possible to find state-of-the-art 
    $c-M$ subhalo relations for the subhalos themselves - by including a radial dependence accounting for the location of the subhalos within the main halo \cite{Moline:2016pbm} - in the following we adopt the one for main halos \cite{Sanchez-Conde:2013yxa}. The main reason for this choice is that the latter is already available in the \texttt{CLUMPY} code, we use to compute the J-factors, while the model by \cite{Moline:2016pbm} is not yet. 
Since the predicted concentrations for subhalos are higher than for main halos of the same mass \cite{Moline:2016pbm}, the parametrization of \cite{Sanchez-Conde:2013yxa} will ensure that our integrated J-factors for the dIrrs will be conservative (by a factor 2-3). This choice also implies to implicitly adopt the standard NFW density profile to model the DM profile in subhalos, $\rho_{clumps}$. However, compared to main halos, subhalos are known to exhibit profiles that are similar to the NFW in the inner parts but decay much more rapidly towards the outskirts due to tidal stripping \cite{Springel:2008cc, Kazantzidis:2003im, Pe_arrubia_2010, errani2020asymptotic}. Yet, for our purposes the NFW still represents a very good approximation, as most ($\sim 90\%$) of the annihilation flux is originated within the scale radius, well inside the subhalo. Once we assume a ($c-M$) relation,  
    there is also the possibility to consider an intrinsic scatter on the concentrations \cite{Bullock:2001jz, Moline:2016pbm}. However, it becomes computationally extremely expensive to take into account this scatter in \texttt{CLUMPY}. Thus, for the purposes of this work, we decide to neglect it: as we are in any case considering very diverse models for the rest of variables, we note that the effect of the scatter in concentrations would lie well within the spread of the obtained J-factors from these different models.
\end{itemize}
After we have discussed and selected the main parameters describing the subhalo population, we will define four different benchmark models for the computation of the J-factors, each of them with a particular level of subhalo boost:
\begin{itemize}
    \item MIN: the main halo is modelled with a Burkert profile and only takes into account the smooth DM distribution within the main halo, i.e. this model neglects the effect of substructure.
    \item MED: the main halo is modelled with a Burkert profile. The SRD follows the Burkert profile of the host. We adopt $\alpha = 1.9$ for the slope of the SHMF (Eq. \ref{eq:subhalo-mass-function}).
    \item MAX-Bur: this model is similar to MED but adopts $\alpha = 2.0$ for the slope of the SHMF.
    \item MAX-NFW: the main halo is modelled with an NFW profile. The SRD follows the NFW profile of the host. We adopt $\alpha = 2.0$ for the slope of the SHMF.
\end{itemize}
We note that with the definition of the above benchmark models, we are bracketing a wide range of different possible substructure scenarios, this way also providing a bracketing for the values of the J-factors. We summarize all these scenarios in Table \ref{tab:benchmark-models}.
Let us stress that, although the labels MIN, MED, MAX already give a rough idea of the ranking of J-factors values, in a few cases this is not strictly true due to the different DM profiles adopted for the main halos in the different models.   

\begin{table}[H]
\centering
\begin{tabular}{ | c | c | c | c | c |}
\hline
Model & $\rho_{host}$ & $\rho_{subs}$ & SRD & $\alpha$ \\
\hline
\rowcolor{lightgray} 
MIN & Burkert & - & - & - \\
MED & Burkert & NFW & Burkert & 1.9 \\ 
\rowcolor{lightgray}
MAX-Bur & Burkert & NFW & Burkert & 2 \\
MAX-NFW & NFW & NFW & NFW & 2 \\ 
\hline
\end{tabular}
\caption{\centering \footnotesize{Summary of the DM models we use in this work; $\alpha$ is the slope in Eq.~(\ref{eq:subhalo-mass-function}), see Section \ref{modeling} for full details.}}
\label{tab:benchmark-models}
\end{table}

The computation of the J-factors is performed using the \texttt{CLUMPY} code \cite{Charbonnier:2012gf, Bonnivard:2015pia, Hutten:2018aix}. \texttt{CLUMPY} allows us to easily implement our four benchmark models for the sample of dIrr galaxies. Other general parameters we use to perform these computations are the minimum subhalo mass, that we set to $M_{min}=10^{-6}$M$_{\odot}$, the maximum subhalo mass in terms of the mass of the host, $M^{\%}_{max}=0.01$, and the number of substructure levels to be considered, $N_{subs}=2$ (subhalos inside subhalos). 
The obtained integrated J-factors for each galaxy and for each benchmark model are summarized in Table \ref{tab:j-factors}.

\begin{table}[H]
\centering
\begin{tabular}{ | c | c | c | c | c | }
\hline
Name & $\log_{10}J_{\rm MIN}$ & $\log_{10}J_{\rm MED}$ & $\log_{10}J_{\rm MAX-BUR}$ & $\log_{10}J_{\rm MAX-NFW}$ \\
\cline{2-5}
& $\rm GeV^2 cm^{-5}$ & $\rm GeV^2 cm^{-5}$ & $\rm GeV^2 cm^{-5}$ & $\rm GeV^2 cm^{-5}$ \\
\hline
\rowcolor{lightgray} 
NGC6822 & 17.86 & 18.40 & 18.62 & 18.63 \\
IC10 & 18.21 & 18.40 & 18.53 & 18.33 \\ 
\rowcolor{lightgray}
WLM & 16.72 & 17.10 & 17.27 & 17.24 \\
IC1613 & 16.16 & 17.48 & 17.74 & 17.84 \\ 
\rowcolor{lightgray}
Phoenix & 14.40 & 15.04 & 15.16 & 15.16 \\
DDO210 & 15.90 & 16.20 & 16.32 & 16.27 \\
\rowcolor{lightgray}
DDO216 & 15.45 & 15.72 & 15.83 & 15.73 \\  
\hline
\end{tabular}
\caption{\centering \footnotesize{Total J-factor values integrated up to $R_{200}$ for each dIrr in our sample, and for the different benchmark models summarized in Table \ref{tab:benchmark-models}.}}
\label{tab:j-factors}
\end{table}

 We point out that the obtained range of J-factors for each dIrr, based on the different substructure scenarios in Table \ref{tab:benchmark-models}, is wider than the one we would have obtained from just having considered the uncertainty on $M_{200}$ (see Table \ref{tab:parameters_bur}) in the J-factor computation for each dIrr. For this reason we neither provide J-factor uncertainties in Table \ref{tab:j-factors} nor include them in our data analysis (see next Section \ref{gammarayDMsection}).

Attending to the obtained values in Table \ref{tab:j-factors}, we note that they are distributed, for all the benchmark models, according to the expected ratio $M_{200}^2/D^2$. From them, we can also easily understand the impact of taking into account different models for the subhalo population inside dIrrs. In particular, we obtain boost values ranging between $B =0.6 - 3.4$ for the MED model, and $B =1.1 - 4.8$ for MAX-Bur, depending on the considered dIrr (we recall that $B=0$ means no boost in our definition). The only exception is IC1613, for which we obtain $B_{\rm MED}=19.9$ and $B_{\rm MAX-Bur}=37.0$. 
For the rest of the objects, we can check how these values compare to the ones obtained in the literature. We can compare with Ref. \cite{Sanchez-Conde:2013yxa}, who also use a $c-M$ parametrization for main halos applied to the subhalos. For masses between $10^{6}-10^{10}$ M$_{\odot}$, they obtain $B =1.2 - 2.0$ for $\alpha=1.9$ and $B =2.0 - 7.0$ for $\alpha=2.0$. This is comparable to our computations, despite the fact that authors in \cite{Sanchez-Conde:2013yxa} implicitly adopted NFW profiles instead of Burkert as we do. The same difference is also found in  
\cite{Bonnivard:2015pia}, where the authors reproduce the same substructure modelling as in \cite{Sanchez-Conde:2013yxa}. They obtain a good agreement, some small differences only arising for high mass values ($M_{200}>10^8$ M$_{\odot}$) and in the case of adopting $\alpha=2$ for the slope of the SHMF. This divergence, that we also find, can be explained by the following: in the case of \cite{Sanchez-Conde:2013yxa}, the total mass in the form of subhalos is added to the original $M_{200}$ of the object, leading to slightly overestimated boost values for higher masses. Instead, in \cite{Bonnivard:2015pia} the mass in the form of subhalos is subtracted from $M_{200}$ of the host, leading to more realistic boost values, as we also obtain. Finally, we can also compare our values to the ones obtained using a $c-M$ relation for subhalos \cite{Moline:2016pbm} (who also adopted NFW profiles for subhalos). In this case, for the same mass range, they obtain $B =2.1 - 3.7$ for $\alpha=1.9$ and $B =3.7 - 10.3$ for $\alpha=2.0$. As expected, these are a factor up to $\sim 2$ higher values, especially for the $\alpha=2.0$ case. The reason being that, as said, is that \cite{Moline:2016pbm} predicts higher concentrations for subhalos when compared to halos. Again, this means that our subhalo-boosted J-factors are conservative. 

\begin{figure}[H]
\centering
\includegraphics[angle=0,height=7.5truecm,width=8.5truecm]{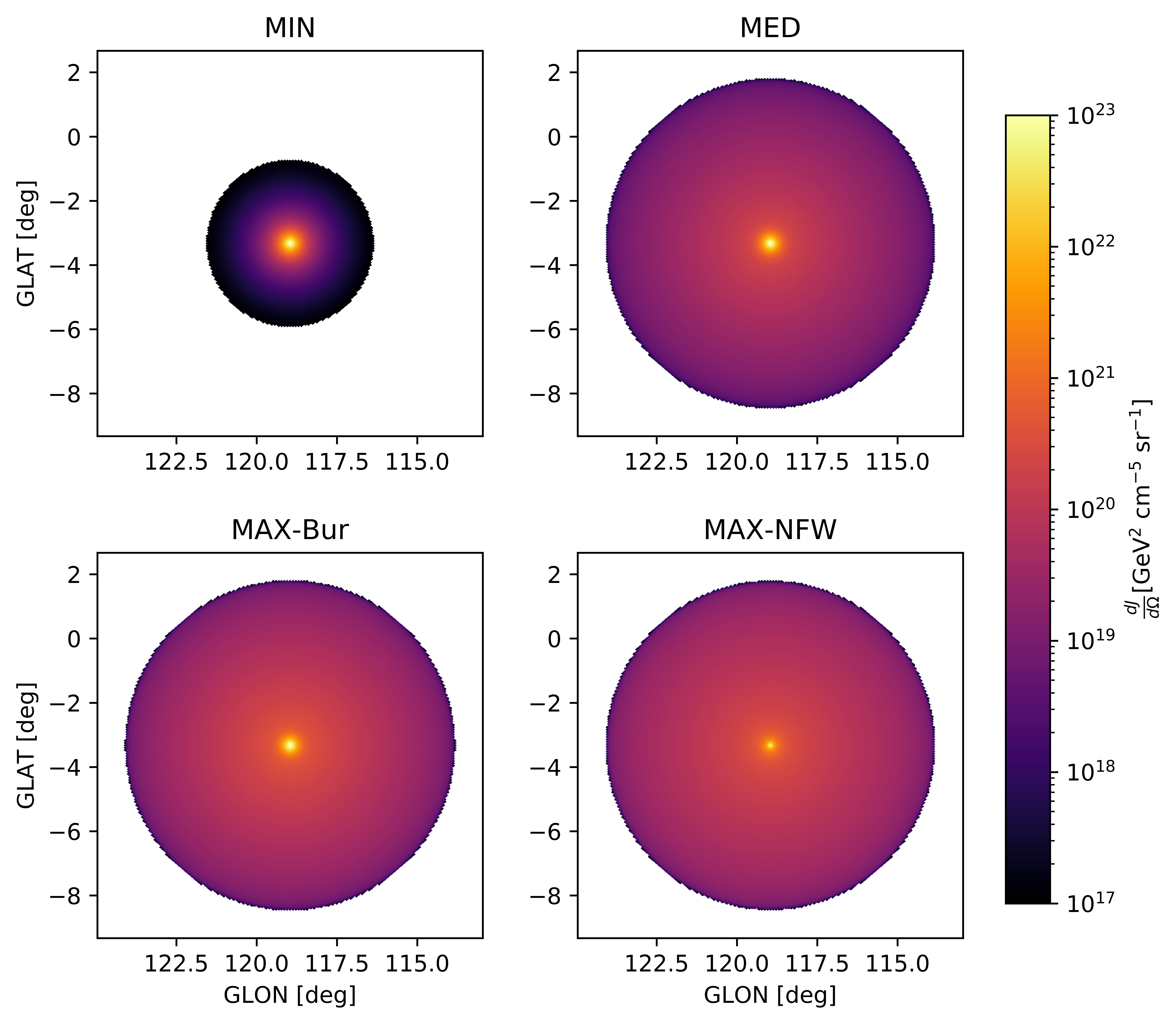}
\caption{\centering \footnotesize{Two-dimensional templates of the expected spatial morphology of the DM annihilation fluxes from the IC10 dIrr, as obtained with \texttt{CLUMPY} for each of the four benchmark DM substructure models in Table \ref{tab:benchmark-models}; see text for details. The z-axis represents the values of the differential J-factors ($\frac{dJ}{d\Omega}$). Because the values of $\frac{dJ}{d\Omega}$ vary by many orders of magnitude in these plots, we chose the colour scale in such a way that we still could appreciate details in the MED and MAX models. However, for the MIN case, this colour scale implies that the outer regions of the object are not visible, as their expected fluxes already lie below the minimum J-factor value shown by the color scale.}}
\label{fig:clumpy_maps_IC10}
\end{figure}

Another output we can easily obtain using \texttt{CLUMPY} are two-dimensional templates, reproducing the spatial morphology of the expected DM annihilation signal. We created maps for each dIrr galaxy and for each benchmark model in Table \ref{tab:benchmark-models}, in total producing a compilation of 28 spatial templates. 
It should be noted that we do not expect any of these subhalos to be individually spatially resolved by \textit{Fermi}-LAT, due to both their extremely small angular extent at the distance of the dIrrs and the LAT instrumental resolution; thus we decided to adopt an averaged description of the whole subhalo population for drawing their contribution to the total signal. An example of these templates is shown in Fig.\ref{fig:clumpy_maps_IC10}. Very importantly, we will use these maps in the next section as the inputs for our \textit{Fermi} spatial and spectral analysis, given that they are the reference models to be fitted to the actual data. 

\section{\textit{Fermi}-LAT data analysis}
\label{gammarayDMsection}
Once the DM modeling of the dIrrs in our sample is complete, we can perform a search for gamma-ray signals in \textit{Fermi}-LAT data. To do so, we use \texttt{Fermipy}, a Python package that automates the ScienceTools\footnote{\url{https://Fermi.gsfc.nasa.gov/ssc/data/analysis/documentation/}} analysis. \texttt{Fermipy v0.19.0} and ScienceTools \texttt{v1.3.7} are used.

The first step is the photon event selection. We will use 11 years of LAT data, from 2008 August 4 to 2019 August 8. The class event is Pass 8 SOURCEVETO \cite{atwood2013pass}, with the corresponding \texttt{P8R3\_SOURCEVETO\_V2} instrumental response function. We choose an energy range from 500 MeV to 1 TeV,\footnote{Although the LAT is sensitive to photons with energies as low as $\sim$20 MeV, the Galactic diffuse emission is much more intense at these energies, and so we decided to start the analysis at higher energies to avoid possible contamination.} with a zenith cut of $\theta_z>105^\circ$.

The data is binned using 8 energy bins per decade in energy and $0.08^\circ$ pixel size, defining a region of interest (ROI) of $12^\circ\times12^\circ$, centered at the position of each dIrr. The Galactic diffuse emission is modeled with the latest LAT template, \texttt{gll\_iem\_v07}, while the isotropic contribution is modeled with the corresponding template, \texttt{iso\_P8R3\_SOURCEVETO\_V2.txt}.

We first perform a baseline fit to each individual ROI using the corresponding \texttt{CLUMPY} template for each of the models. The spectral energy distribution (SED) parameters of all the sources in the ROI, the normalization of the Galactic diffuse emission and the isotropic template are left free. As the analysis uses 3 years more data than the 4FGL, the pipeline searches for new sources. These eventual sources would then be added to the model to optimize the fit of the ROI; in this analysis, no new sources have been found.

Then, we run a fit and compute the likelihood profile as a function of energy and energy flux of DM. In each energy bin, the only free parameter is the normalization, which is computed independently from other bins.\footnote{By analyzing each energy bin separately, we avoid selecting a single spectral shape to span the entire energy range, at the expense of introducing additional degrees of freedom in the fit.} We then scan for each energy bin the likelihood as a function of the flux normalization for the assumed DM signal, which depends both on the annihilation channel and the WIMP mass, and adopt those parameters which maximize the likelihood.

In the following, we will consider three annihilation channels: $b\bar{b}$, $\tau^+\tau^-$, and $W^+W^-$, and WIMP masses ranging from 5 GeV up to 10 TeV. For each channel and mass, we extract the expected flux, $\phi_{\gamma,j}$, in each energy bin. The upper limits to the flux for every bin and target are shown in Appendix \ref{app:flux_upper_limits}. Then, we compute the likelihood of observing $\phi_{\gamma,j}$, and the log-likelihood in each energy bin is summed to get the overall log-likelihood, given by,

\begin{equation}
    \mathrm{log}\mathcal{L}\left(\mu,\theta|\mathcal{D}\right)=\sum_j \mathrm{log}\mathcal{L}_j\left(\mu,\theta_j|\mathcal{D}_j\right)
\end{equation}

\noindent where $\mathcal{L}$ is the likelihood, $j$ is the index of each energy bin of the \textit{Fermi}-LAT data ($\mathcal{D}$), $\mu$ are the DM parameters ($\langle\sigma v\rangle$ and $m_\chi$), and $\theta$ are the parameters in the background model, i.e., the nuisance parameters. These include the uncertainty of the J-factors, set to 0.3 dex (i.e., $\mathrm{log}_{10}~\sigma_J=0.3$) in our analysis for every target in the sample. This value was chosen following the typical size of J-factor uncertainties reported in the literature \cite{Bonnivard:2015pia,chiappo+15,PaceAndStrigari19} as well as previous choices by the LAT collaboration \cite{LATPass8dwarfs,DrlicaWagner_LATDESdwarfs16,Albert_2017}, associated with systematic uncertainties in the determination of the profile parameters, i.e. the fit of the RC.\footnote{
 The uncertainty on individual J-factor values is subdominant compared to the one coming from the use of different DM models for each dwarfs, i.e., MIN, MED and MAX models. Nonetheless, we decided to include it in our analysis even if this choice will not affect the results, being the systematics associated to the different DM models dominant.} The significance of the DM hypothesis can be evaluated via the test statistics (TS)\footnote{$TS\sim\sigma^2$, this is, a $5\sigma$ detection would be equivalent to $TS\sim25$. Strictly speaking, $TS\sim\sigma^2$ only applies in the asymptotic case with nested hypotheses and 1 additional degree of freedom. Furthermore, the null hypothesis can't be degenerate.},

\begin{equation}
    TS=2~\Delta\mathrm{log}\mathcal{L}=2~\mathrm{log}\left[\frac{\mathcal{L}\left(\mu,\theta|\mathcal{D}\right)}{\mathcal{L}_{null}\left(\theta|\mathcal{D}\right)}\right]
\end{equation}

\noindent where $\mathcal{L}_{null}$ is the likelihood in the case of null hypothesis, i.e., no DM, and $\mathcal{L}$ is the likelihood for the DM hypothesis.\\

\begin{figure}[ht!]
    \centering
    \includegraphics[angle=0,height=5.5truecm,width=7.5truecm]{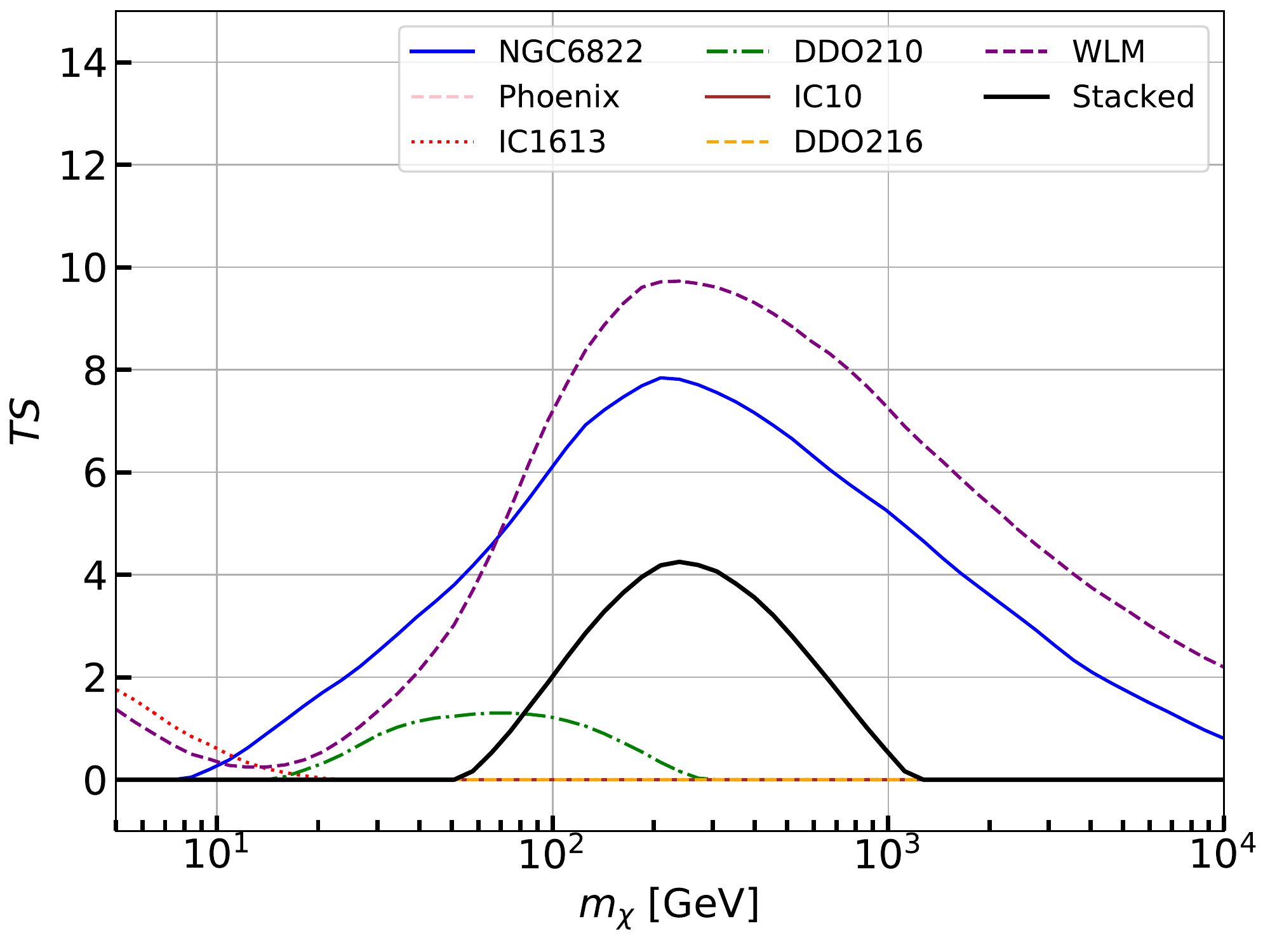}
    \includegraphics[angle=0,height=5.5truecm,width=7.5truecm]{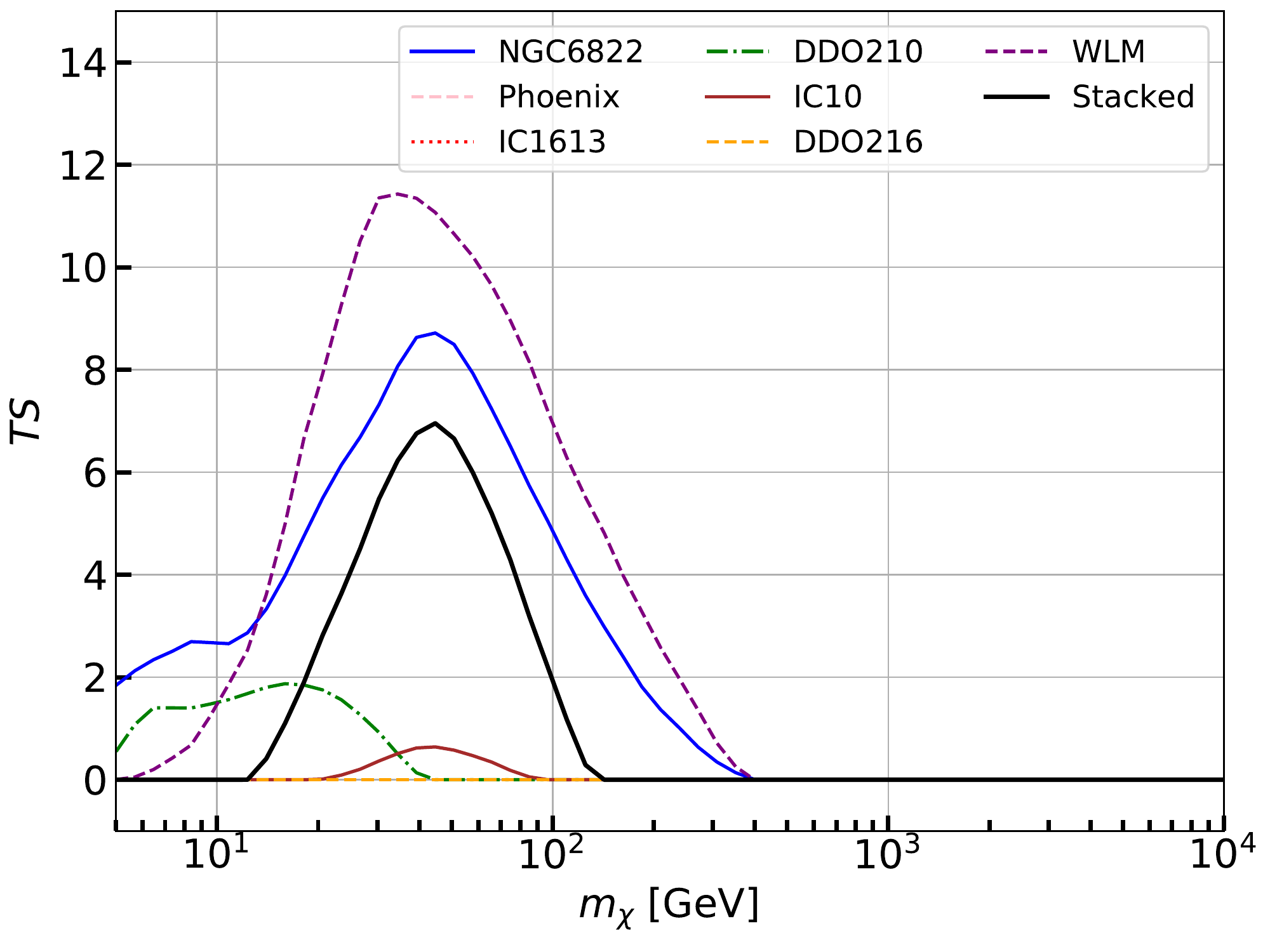}
    \includegraphics[angle=0,height=5.5truecm,width=7.5truecm]{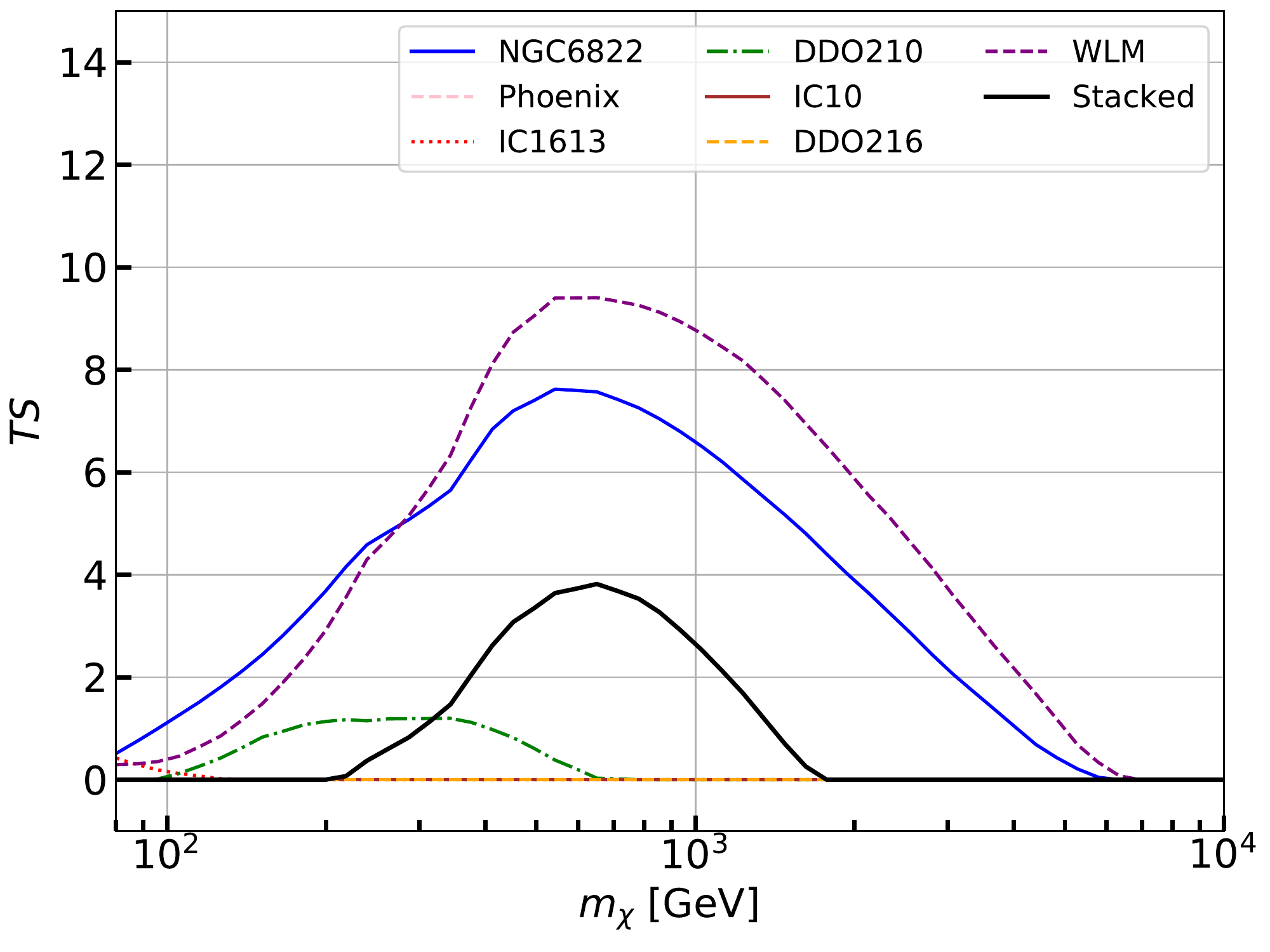}
    \caption{\centering \footnotesize{Likelihood profiles as a function of the WIMP mass, for each of the dIrrs and the combined targets, assuming the MED model. The $\langle\sigma v\rangle$ is left free for each mass bin. Top, middle and bottom panels are for the $b\bar{b}$, $\tau^+\tau^-$, and $W^+W^-$ annihilation channels, respectively.
    }}
    \label{fig:likelihood_profiles}
\end{figure}

Once the individual targets have been studied, one can perform a combined analysis simultaneously using all the targets in our sample. This is performed by simply summing the individual log-likelihood profiles for each of the targets, to obtain a global likelihood. 
As seen in Figure \ref{fig:likelihood_profiles}, where the MED model is used, no significant emission is detected from any of the targets and any of the channels, with the WLM galaxy having the largest observed TS, $\sim$9--11, depending on the annihilation channel. We also show the combined likelihood when considering a joint analysis. It is interesting to look at the likelihood profile of each dIrr as a function of $m_\chi$, i.e., the TS preference for each of the masses we are scanning. The wider the profile, the more uncertain is the determination of the potential signal, while the height measures the overall preference.

Note that these likelihoods present each mass bin with its own best-fit $\langle\sigma v\rangle$. From Figure \ref{fig:likelihood_profiles}, a $\sim 3\sigma$ excess is seen in the cases of WLM and NGC6822. This apparent excess is present in the three considered annihilation channels, while its position shifts to slightly larger WIMP masses in the case of $W^+W^-$ and lower ones in $\tau^+\tau^-$. The shift is just due to the fact that the $b\bar{b}$ channel peaks at roughly $E_{peak}
^{b\bar{b}}\sim m_\chi/20$, while $E_{peak}
^{W^+W^-}\sim m_\chi/30$ and $E_{peak}
^{\tau^+\tau^-}\sim m_\chi/3$. 
It is interesting to note that both the excesses in WLM and NGC6822 peak at the same masses (as expected from an universal DM signal), and that in the $b\bar{b}$ channel the peak is at $m_\chi\sim250$ GeV, which is still allowed by the DM constraints obtained from the dSphs~\cite{oakes2019combined}. Yet, these excesses are most likely due to the Galactic diffuse emission, which is not perfectly modeled in any LAT analysis 
\footnote{We note that authors in Ref.~\cite{Gammaldi:2017mio} computed the expected astrophysical gamma-ray emission {\it originated in} these objects to be well below the LAT detection threshold.}. In any case, these TS values are not considered to be significant, as they are pre-trials and therefore are expected to decrease significantly in a more complete statistical analysis including the ``look-elsewhere'' effect. Thus, as no gamma-ray emission is conclusively observed from any of the targets, in the next subsection we will proceed to set limits to the WIMP mass vs. annihilation cross section parameter space.

\section{Dark matter limits on the annihilation cross section}
\label{sec:DMlimits}
\begin{figure}[ht!]
\centering
\includegraphics[angle=0,width=0.75\linewidth]{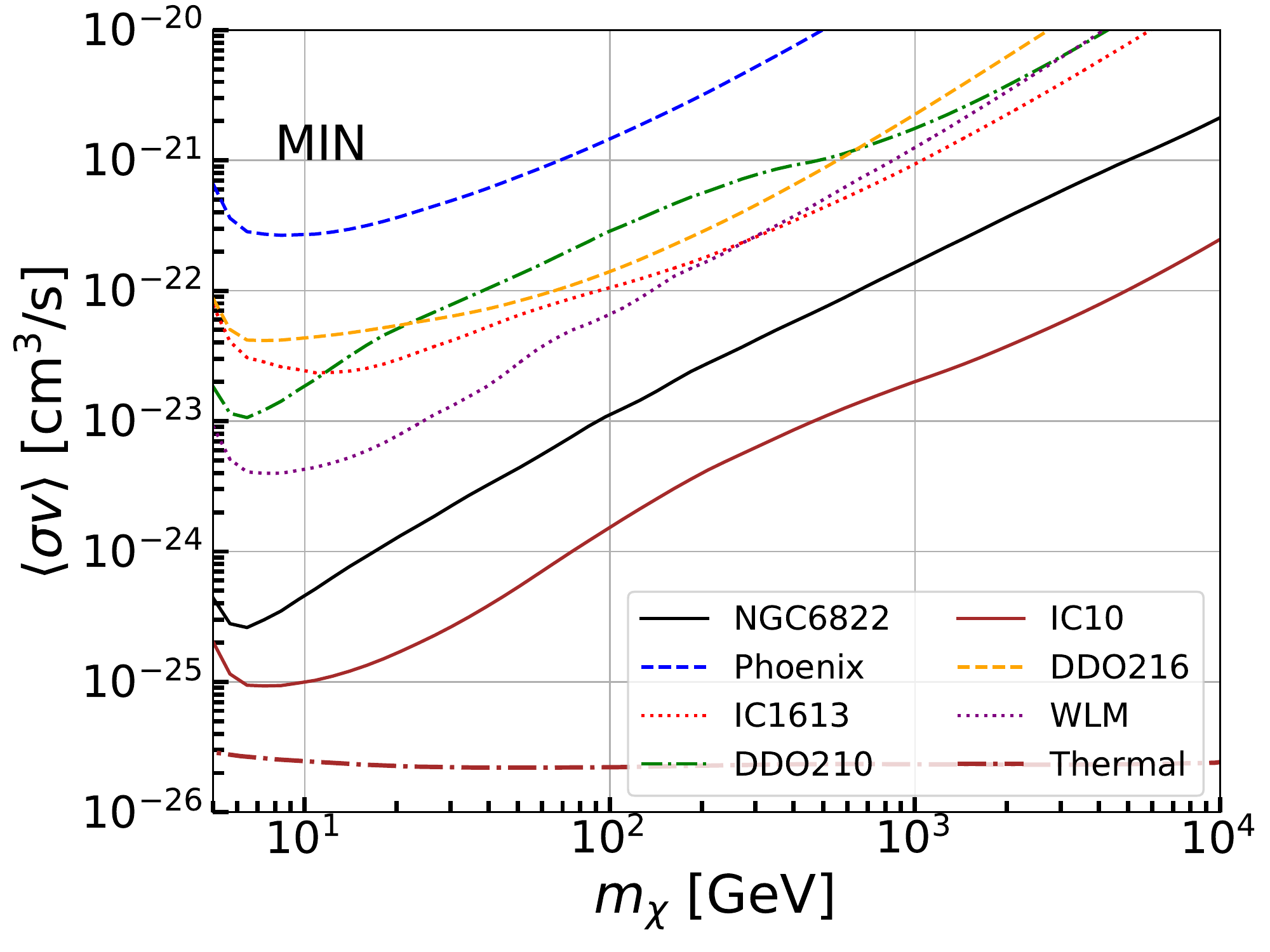}
\includegraphics[angle=0,width=0.75\linewidth]{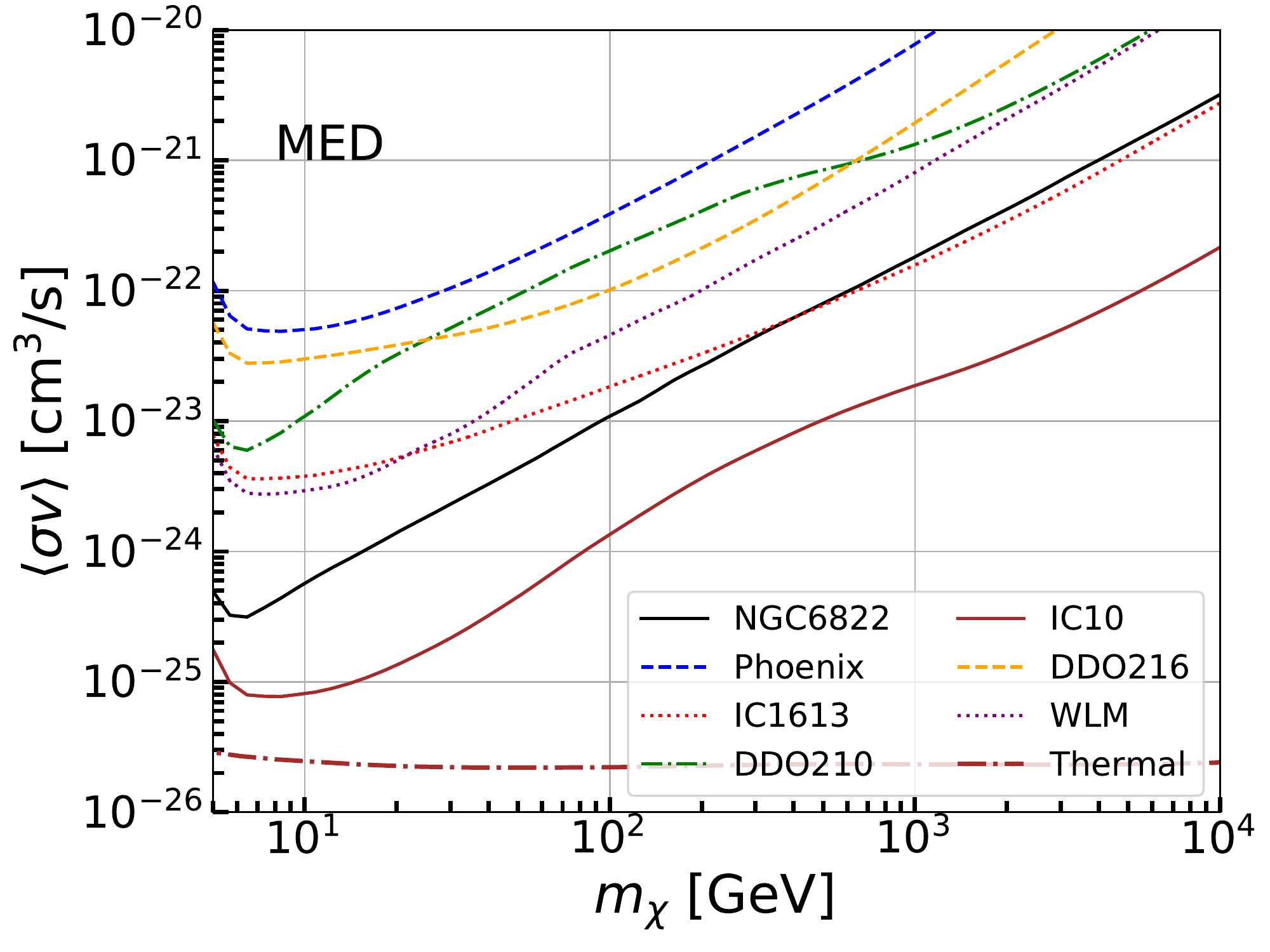}
\includegraphics[angle=0,width=0.75\linewidth]{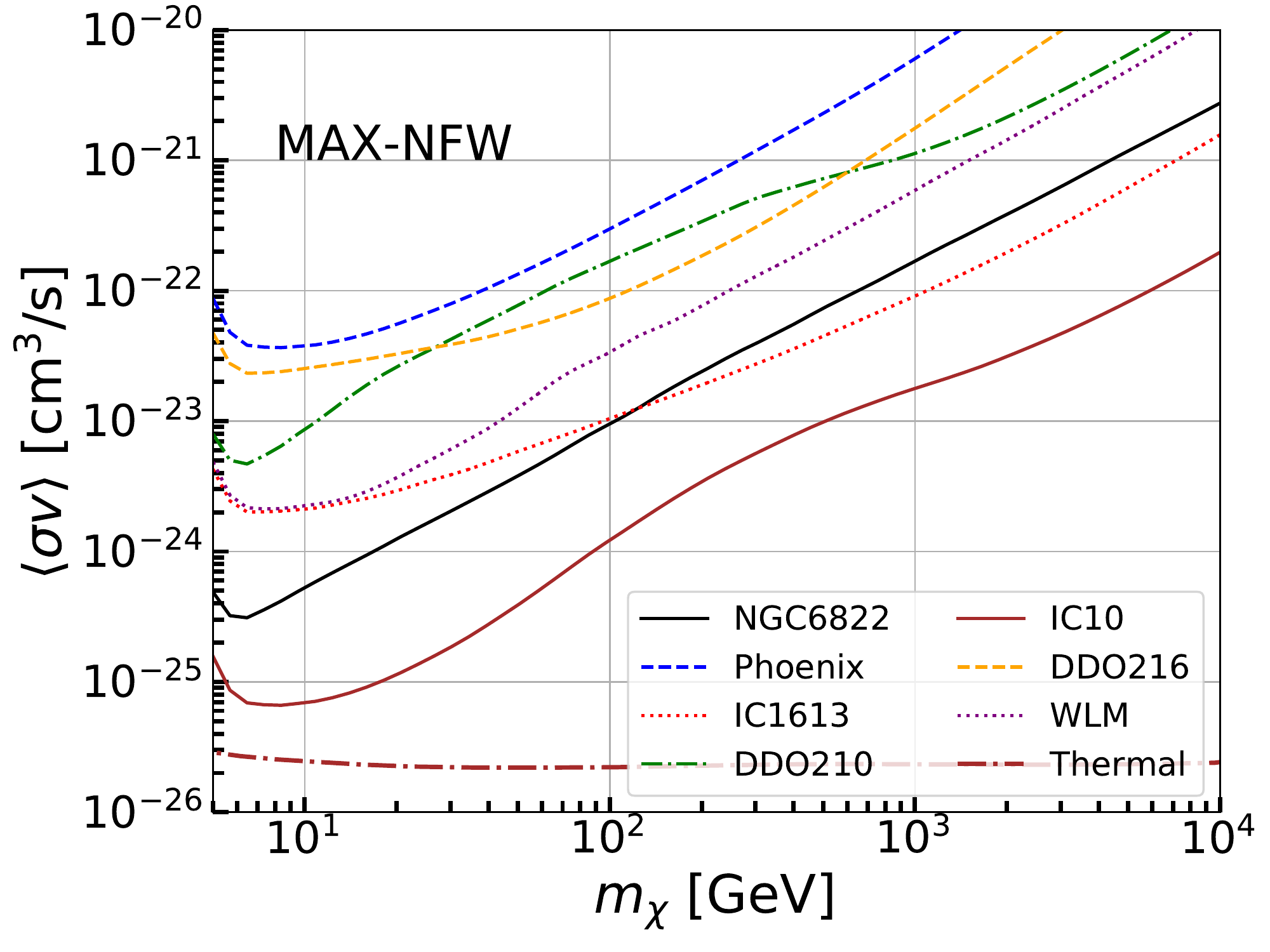}
\includegraphics[angle=0,width=0.75\linewidth]{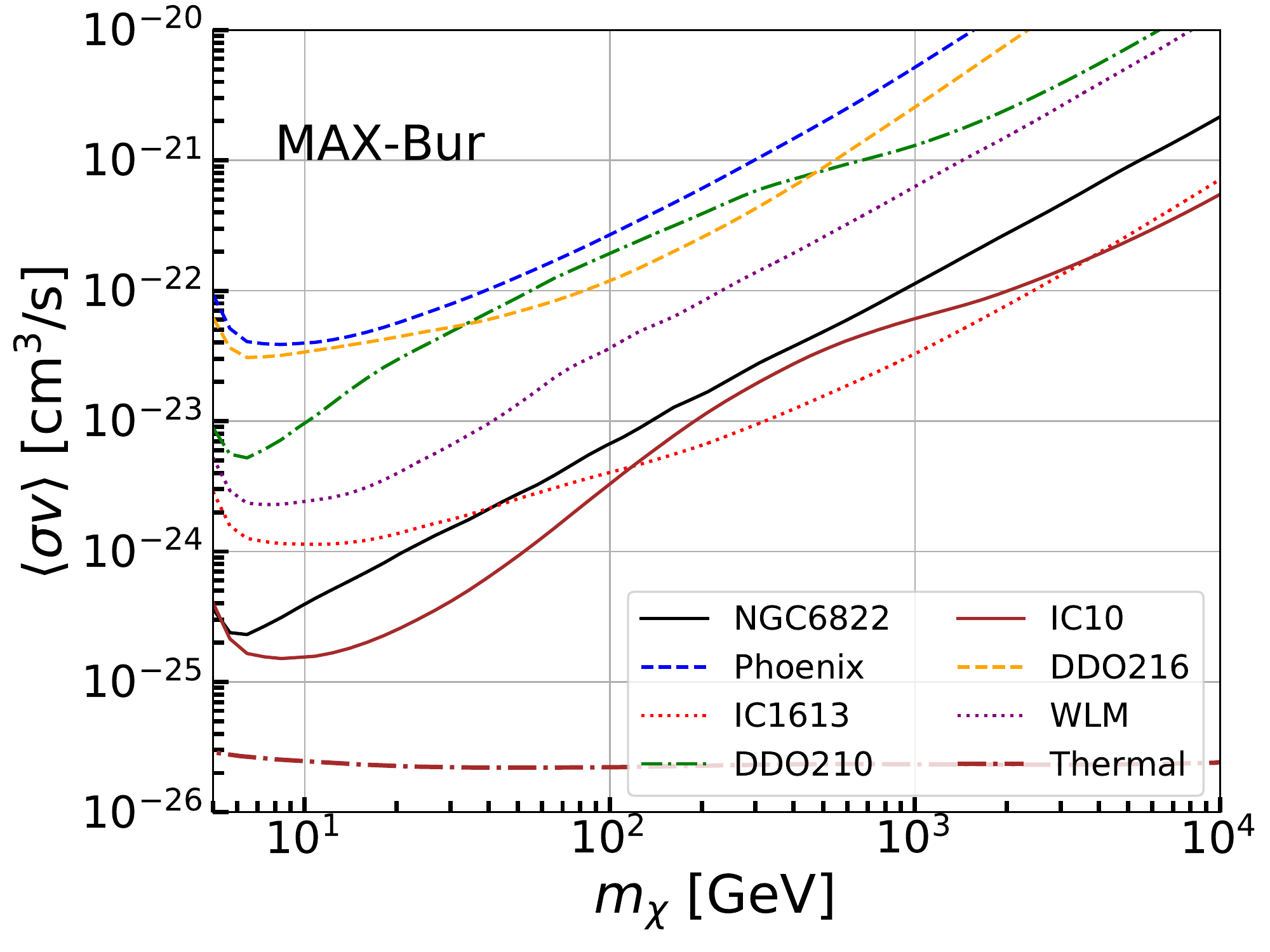}
\caption{\centering \footnotesize{Upper limits for $\langle\sigma v\rangle$ for each individual source, for the $b\bar{b}$ annihilation channel. The different DM models considered are, from top to bottom, MIN, MED, MAX-Bur and MAX-NFW; see Table \ref{tab:benchmark-models} for details on each of them. 
}}
\label{fig:bb_individual}
\end{figure}

From $\mathcal{L}\left(\mu,\theta|\mathcal{D}\right)$ we can evaluate the one-sided 95\% confidence level (CL) exclusion limit on the flux, 
which is the value at which the log-likelihood decreases by 2.71 (as we consider one-sided limits) with respect to its maximum value. Then, from this value and Eq.~(\ref{eq:dm-flux}) we can compute 95\% C.L. upper limits in the $\langle\sigma v\rangle-m_\chi$ parameter space for each dIrr and DM modeling scenario described in Section \ref{modeling}. In Figure \ref{fig:bb_individual} the individual limits are plotted for the four considered DM models and the $b\bar{b}$ annihilation channel. The individual limits for $\tau^+\tau^-$ and $W^+W^-$ are deferred to Appendix \ref{app:individual_limits_tt_ww}.

As seen in the figure, none of the targets is able to reach the canonical, thermal relic cross section, the best limits from IC10 still being a factor $\sim$10 away. Indeed, the best results are obtained for IC10 in all DM models which, interestingly, have almost no sensitivity to the change of the DM density models, as the four considered models yield very similar results. This is not true for the rest of dIrrs though, for which the change of DM model can vary the limits up to a factor 10 (e.g., IC1613, for which a significantly different behavior of its extended emission under the different DM scenarios was obtained).

\begin{figure}[ht!]
\centering
\includegraphics[angle=0,height=6truecm,width=8truecm]{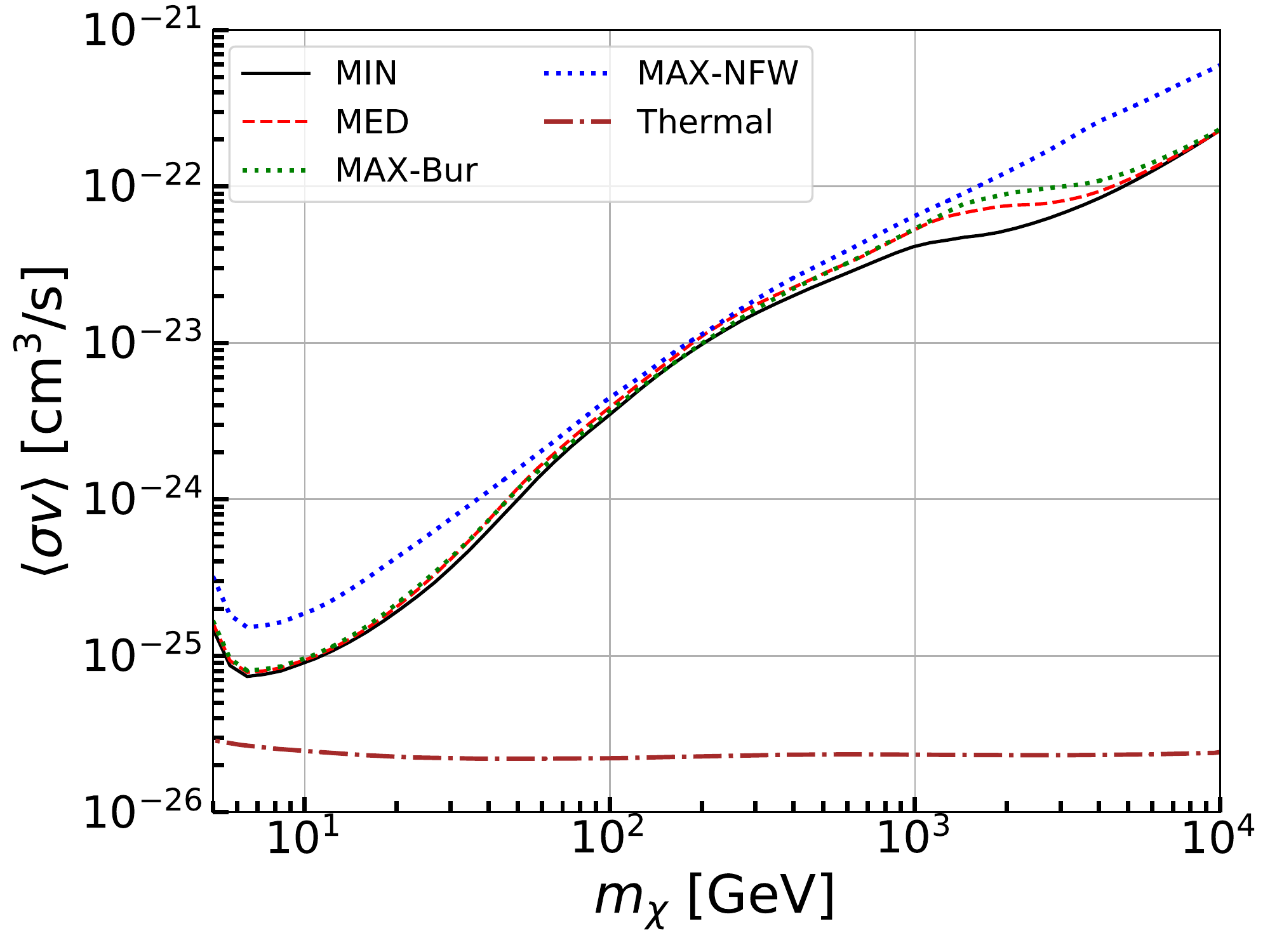}
\includegraphics[angle=0,height=6truecm,width=8truecm]{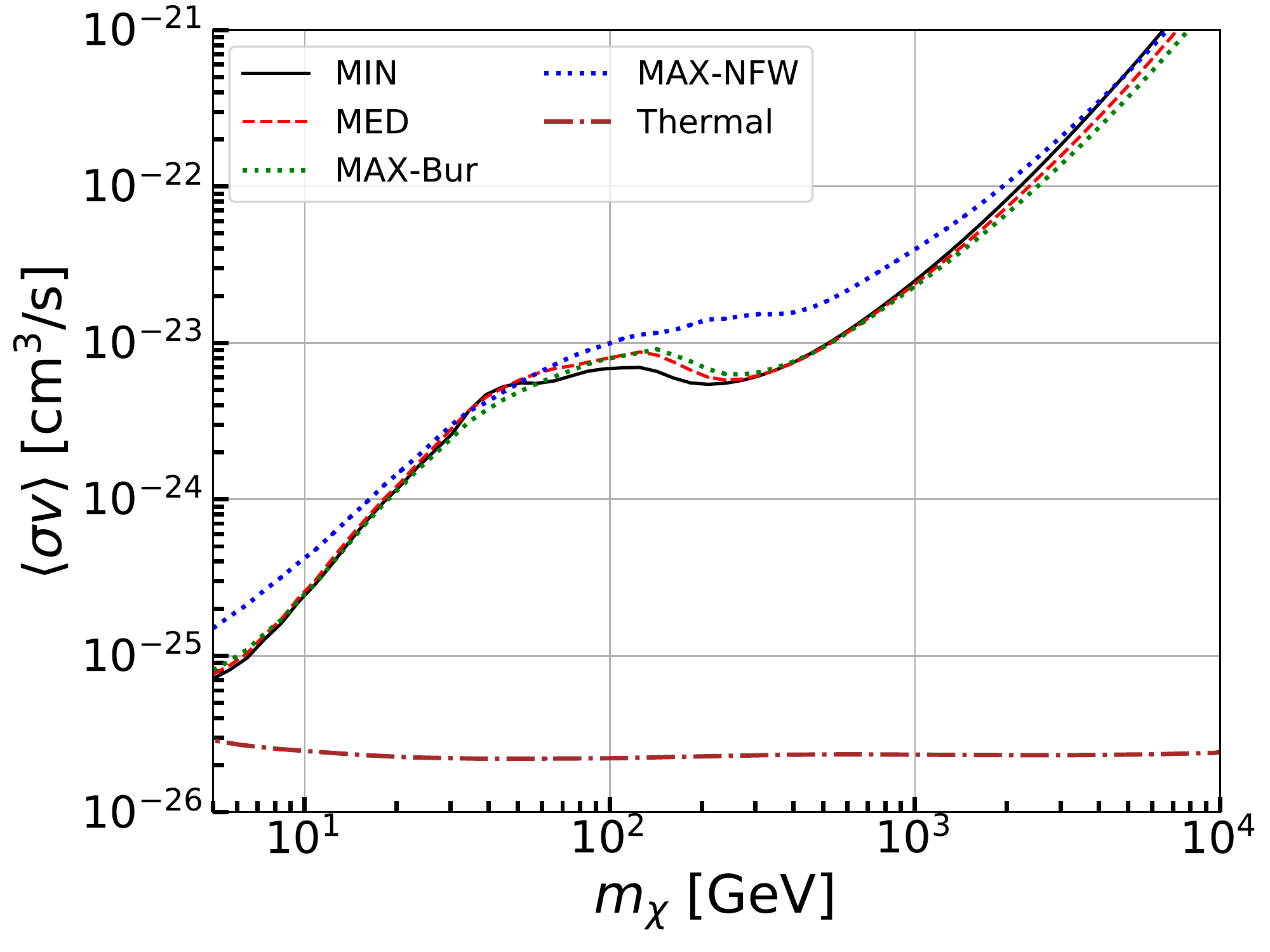}
\includegraphics[angle=0,height=6truecm,width=8truecm]{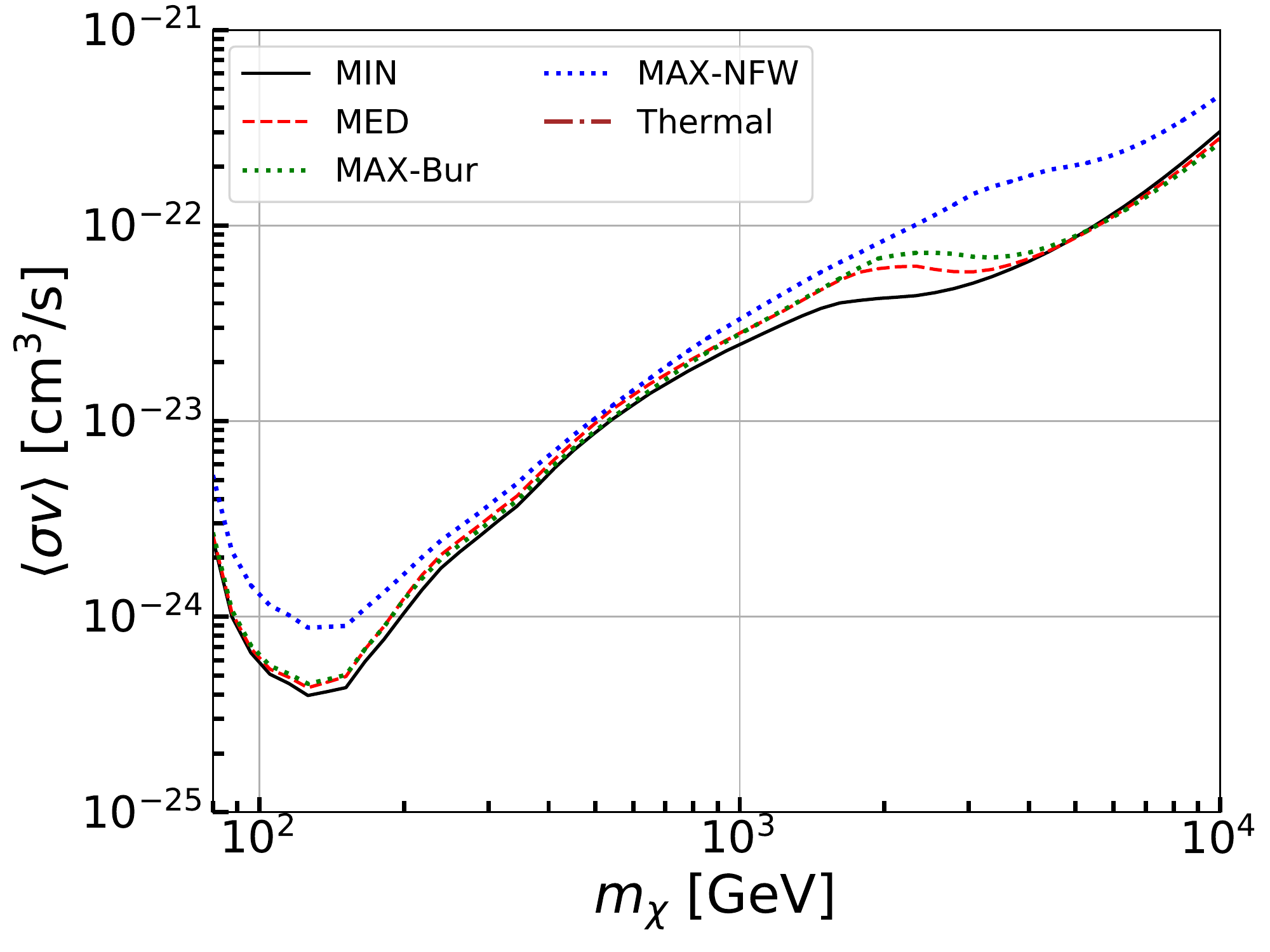}
\caption{\centering \footnotesize{Limits for the combined signal in the four considered scenarios, for $b\bar{b}$ (top panel), $\tau^+\tau^-$ (middle panel) and $W^+W^-$ (bottom panel) annihilation channels. 
}}
\label{fig:stacked_limits}
\end{figure}

Our joint likelihood analysis allows also to derive combined DM constraints using all objects in the sample at once. These constraints are plotted in Figure \ref{fig:stacked_limits} for the four considered DM models and the three annihilation channels.

The combined limits are dominated by the brightest (i.e., with largest J-factor) object, IC10, whose limits we recall only changed marginally between the different DM models. Therefore, the combined limits present a very small deviation from each other as well, and for the $b\bar{b}$ channel are at the level of $\mathrm{\sim10^{-25}cm^3s^{-1}}$ for $m_\chi=10$ GeV and $\mathrm{\sim5\cdot10^{-23}cm^3s^{-1}}$ for $m_\chi=1$ TeV.

We note that the observed local excesses in the case of NGC6822 and WLM (see Figure \ref{fig:likelihood_profiles}) weaken the joint limits at the corresponding masses in each annihilation channel. As the combined analysis is dominated by the brightest objects, WLM has the most relevant contribution to this weakening, which occurs for the masses at which the excess has the largest TS values, i.e, around 200--300 GeV for $b\bar{b}$, 30--70 GeV for $\tau^+\tau^-$ and 500--600 GeV for $W^+W^-$.

\section{Discussion and Conclusions}
\label{conclusions}

We analyzed 11 years of \textit{Fermi}-LAT data from the sky regions corresponding to 7 dwarf irregular galaxies, i.e. NGC6822, IC10, WLM, IC1613, Phoenix, DDO210 and DDO216 in the context of DM searches. DIrrs are rotationally-supported star-forming galaxies, yet DM dominated systems, thus suitable targets for indirect DM searches. 
Nevertheless, they represent a clear example of the cusp-core tension between observations and N-body simulations.  
For this reason, in our work we considered both a data-driven core-like Burkert and an N-body simulation motivated cuspy NFW DM density profile. 
\\
We used \texttt{CLUMPY} in order to calculate the J-factors for each target and DM profile, which include the effect of subhalos in the annihilation flux under different configurations of the subhalo population in these objects. 
The values obtained for the subhalo boost for our benchmark DM models reach a factor $\sim 5$, i.e. a factor $\sim 2$ lower than the ones obtained in \cite{Moline:2016pbm} for this same mass scale. This means that we derived both conservative J-factors and DM constraints. For each DM model, we created two-dimensional spatial templates of the expected DM annihilation signal with \texttt{CLUMPY}. In fact, the angular extension of dIrr galaxies 
makes it mandatory to consider it in the data analysis should we want it to be state-of-the-art and realistic. \\ 
We have performed a search for gamma-ray signals in \textit{Fermi}-LAT data in each of the targets' ROI. 
After our analysis, these objects stay undetected in gamma-rays. 
No  significant  emission  is  detected, with the highest TS values $TS\sim 9-11$ corresponding to the WLM galaxy ($TS\sim 8$ for NGC6822), depending on the considered annihilation channel ($b \bar b$, $\tau^+\tau^-$ and $W^+W^-$).  These TS values are not considered to be significant, mainly because they are pre-trials and thus are expected to  decrease  significantly in a more complete statistical analysis. Also, the presence of little excesses at a few GeV is common in this type of analysis due to our imperfect knowledge of the Galactic foregrounds, which may contribute at this TS level. Nevertheless, this fact could point to a potential DM emission in the two objects. Indeed, based on previous theoretical studies, we expect a negligible astrophysical flux from SFRs in these galaxies and the DM-induced emission should be the dominant one \citep{Gammaldi:2017mio}.\footnote{Assuming the significance scales as the square root of time in this energy regime, many additional years of LAT data would still be required to confirm a signal, if such a signal is actually present.}\\
Since no gamma-ray emission is conclusively observed from any of the targets, 
we use our flux upper limits to set constraints on the WIMP mass vs. annihilation cross section parameter space. 
The most stringent constraints are obtained for IC10 and NGC6822, independently of the adopted DM profile, and are at the level of $<\sigma v> \sim 10^{-25}-10^{-22}\, \text{cm}^3\text{s}^{-1}$ for $m_\chi \sim 10 - 10^{4}$ GeV, respectively. 
Differences between limits are mainly due to the different contribution that the halo substructure boost has for each object in the sample. 
Finally, we obtained combined DM limits from the joint likelihood data analysis performed. 
IC10 also dominates these combined limits 
for each model and three annihilation channels, i.e. $b \bar b$, $\tau^+\tau^-$ and $W^+W^-$. The strongest constraints are obtained for the $b \bar b$ annihilation channel and are at the level of $<\sigma v>\sim 7\times 10^{-26}\,\text{cm}^{3}\text{s}^{-1}$ at $m_\chi \sim 6$ GeV.\\
\\

\begin{figure}[t!]
    \centering
    \includegraphics[angle=0, width=1\linewidth]{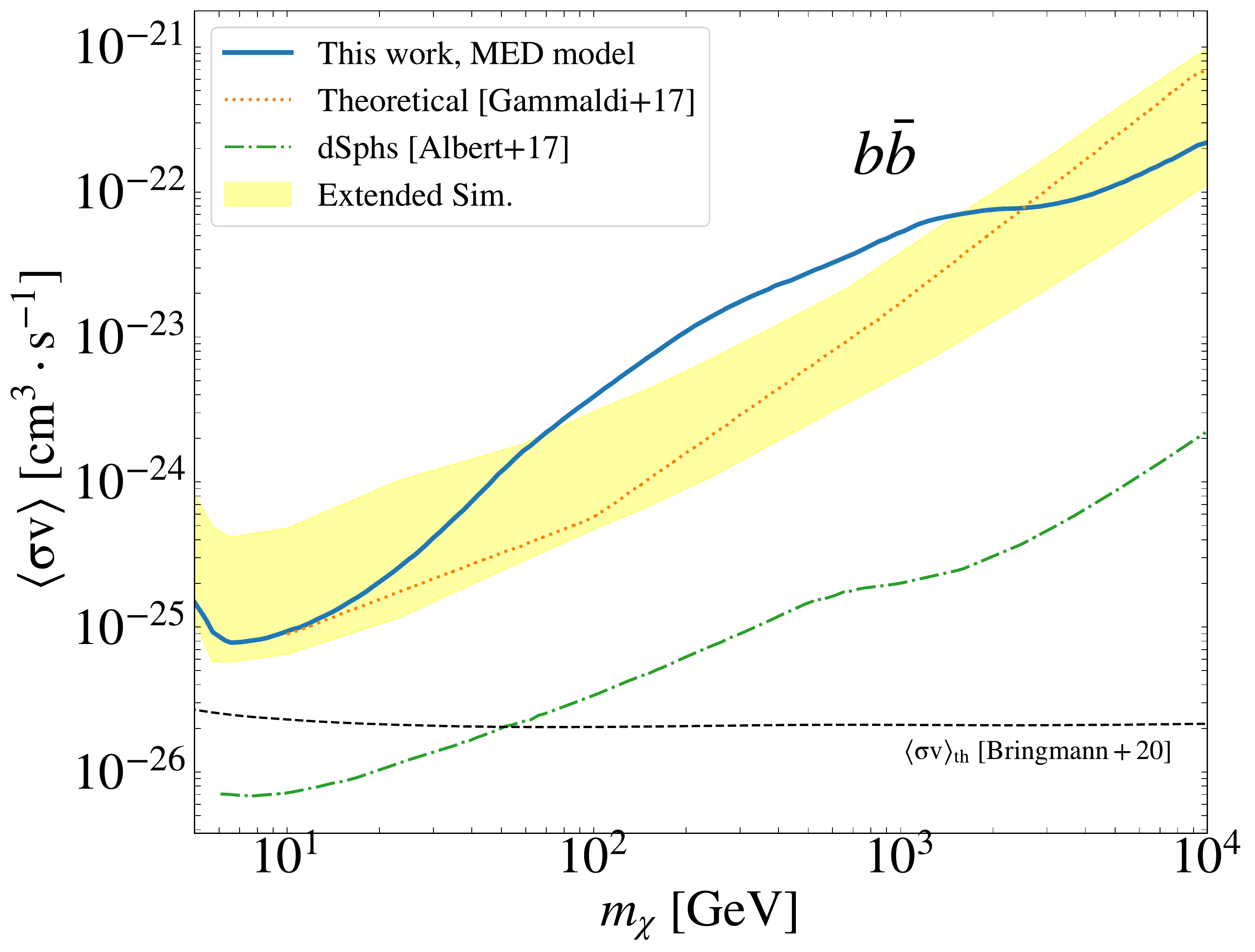}
    \caption{\centering \footnotesize{Comparison between different limits for the $b\bar{b}$ annihilation channel. We include the ones derived in this work, assuming the spatial template (solid blue), the theoretical predictions from \cite{Gammaldi:2017mio} (dashed orange), the null extended simulations 95\% containment band (see Appendix \ref{app:simulations}), and the LAT dSphs \cite{Albert_2017} in dot-dashed green.}}
    \label{fig:comparison}
\end{figure}
In Figure \ref{fig:comparison} we showed the main results of this work: the combined DM limits obtained from the spatial data analysis 
for the MED model and the $b \bar b$ annihilation channel (blue line). We note that these limits are slightly different from the yellow band shown in the same plot, which represents the $95\%$ C.L. containment band after having performed 100 control simulations assuming no DM content in the targets, yet modeled with their corresponding spatial templates (null simulations; see App. \ref{app:simulations} for further details). The observed mismatch can be easily attributed to the small local TS excesses found for some objects in our sample at the relevant energies. 

Interestingly, our combined DM limits are in remarkable agreement with the constraints obtained in the previous theoretical work (yellow-dotted line) \cite{Gammaldi:2017mio}, where the Universal Rotation Curve was assumed as the basis for the DM modelling and the targets were considered to be point-like just as a first approximation. For comparison, in Figure \ref{fig:comparison} we also show the constraints obtained by the \textit{Fermi}-LAT collaboration from the combined analysis of tens of dSph galaxies \cite{Albert_2017}. 
The latter allows to rule out light thermal WIMP masses below $\sim$60 GeV \cite{Calore:2018sdx}. In contrast, the limits from our combined analysis are not able to reach the canonical, thermal relic cross section. The best limits are reached at the lightest considered WIMP masses, still being a factor $\sim$3 above the thermal value. \\
 Let us stress that our work represents the first extended analysis of dIrr galaxies with 11 years of \textit{Fermi}-LAT data in the context of DM searches. By collecting more and more spectral data and RC measurements, it will be possible to obtain a better estimation of the DM halo mass associated to these objects, in this way further reducing the corresponding uncertainty in this type of analysis. Also the use of future, more refined models for both the DM density profile and the annihilation boost due to substructures will help in this direction. Certainly, a better understanding of baryonic physics by means of hydrodynamical simulations and its comparison with the available observational data \cite{2014MNRAS.439.1015K} would also help in order to reach a complete understanding of the kinematics of these objects.\\
 To conclude, the increasing number of dIrr galaxies that has been recently detected 
and studied in their kinematics (see e.g. \cite{Oh:2015xoa, oh1, 2020ApJ...898..102Y, 2020MNRAS.498.5885B}) , make them interesting targets for gamma-ray DM searches and further efforts should be pursued to the study of this class of dwarfs. 
Among others, the Study of H$\alpha$ from Dwarf Emissions (SH$\alpha$DE) \cite{2020MNRAS.498.5885B} is a high spectral resolution integral field survey of 69 dwarf galaxies\footnote{49 star-forming galaxies selected form the Sloan Digital Sky Survey Data Release \cite{2015ApJS..219...12A} as well as 20 targets from the SAMI survey \cite{10.1111/j.1365-2966.2011.20365.x} as a control sample.} with stellar masses $10^6<M_D<10^9 \,\rm{M_\odot}$. 
SH$\alpha$DE is designed to study the kinematics and stellar populations of dwarf galaxies using consistent methods applied to massive galaxies and at matching level of detail, connecting these mass ranges in an unbiased way. 
\\
\\


\vspace{1cm}
\begin{acknowledgments}
The work of VG, JCB, JPR and MASC was supported by the Spanish Agencia Estatal de Investigaci\'on through the grants PGC2018-095161-B-I00 and IFT Centro de Excelencia Severo Ochoa SEV-2016-0597, the {\it Atracci\'on de Talento} contract no. 2016-T1/TIC-1542 granted by the Comunidad de Madrid in Spain, and the MultiDark Consolider Network FPA2017-90566-REDC. VG's contribution to this work has been supported by Juan de la Cierva-Formaci\'on FJCI-2016-29213 and Juan de la Cierva-Incorporaci\'on IJC2019-040315-I grants. JPR also acknowledges the support by MINECO via the SEV-2016-0597-17-2 grant, financed jointly with Fondo Social Europeo. EK work was supported by the grant ``AstroCeNT: Particle Astrophysics Science and Technology Centre" carried out within the International Research Agendas programme of the Foundation for Polish Science financed by the European Union under the European Regional Development Fund. \\
The \textit{Fermi} LAT Collaboration acknowledges generous ongoing support
from a number of agencies and institutes that have supported both the
development and the operation of the LAT as well as scientific data analysis.
These include the National Aeronautics and Space Administration and the
Department of Energy in the United States, the Commissariat \`a l'Energie Atomique
and the Centre National de la Recherche Scientifique / Institut National de Physique
Nucl\'eaire et de Physique des Particules in France, the Agenzia Spaziale Italiana
and the Istituto Nazionale di Fisica Nucleare in Italy, the Ministry of Education,
Culture, Sports, Science and Technology (MEXT), High Energy Accelerator Research
Organization (KEK) and Japan Aerospace Exploration Agency (JAXA) in Japan, and
the K.~A.~Wallenberg Foundation, the Swedish Research Council and the
Swedish National Space Board in Sweden. Additional support for science analysis during the operations phase is gratefully
acknowledged from the Istituto Nazionale di Astrofisica in Italy and the Centre
National d'\'Etudes Spatiales in France. This work performed in part under DOE
Contract DE-AC02-76SF00515.

\end{acknowledgments}

\bibliography{biblio}

\include{PRD_Fermi_v1.bbl}


\appendix
\renewcommand\thefigure{\thesection.\arabic{figure}}    
\section{Rotation curves }
\label{AppA}

In this Appendix, we show the fits to the rotation curves adopting a Burkert DM density profile; see Figure~\ref{RCs}. For all panels, green solid lines in the zoom-in regions show the total fit, i.e., DM component (Burkert profile) and baryons (gas plus stars in the disk). Blue solid lines show the DM-only, Burkert component. We also show as blue dashed lines the result of reconstructing, for each case, the rotation curves with the $\Lambda$CDM consistent (DM-only) NFW profile introduced in section~\ref{Irr} with some assumptions. The main discrepancies between the model and the data for some objects may be associated with (i) an incomplete knowledge of baryonic effects in numerical simulations and/or (ii) systematic uncertainty in the determination of the RC data, e.g., the estimated inclination angle of the galaxy. (see section~\ref{Irr} for further details). 

\begin{figure*}
    \centering
    \includegraphics[angle=0,width=0.4\linewidth]{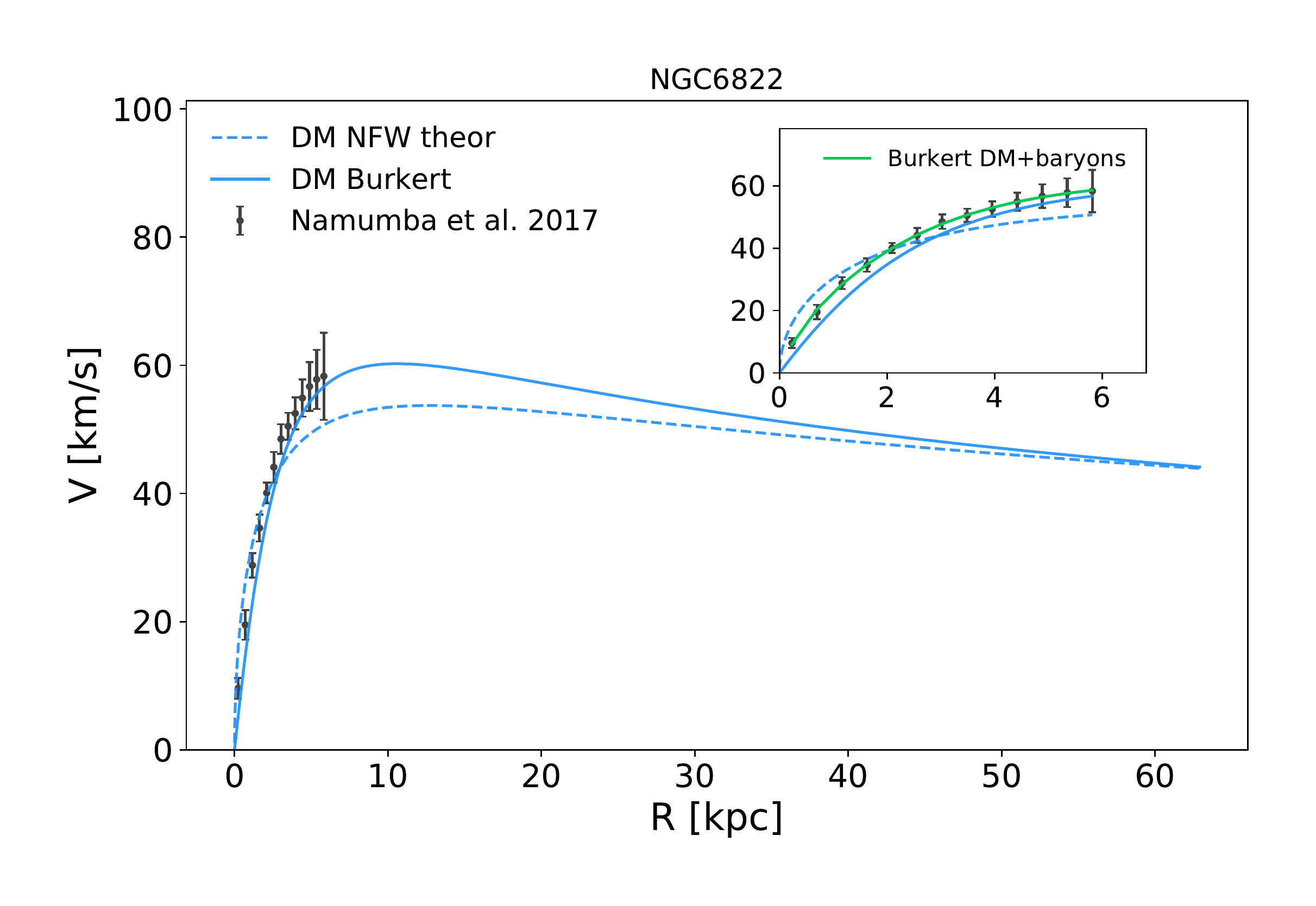}
    \includegraphics[angle=0,width=0.4\linewidth]{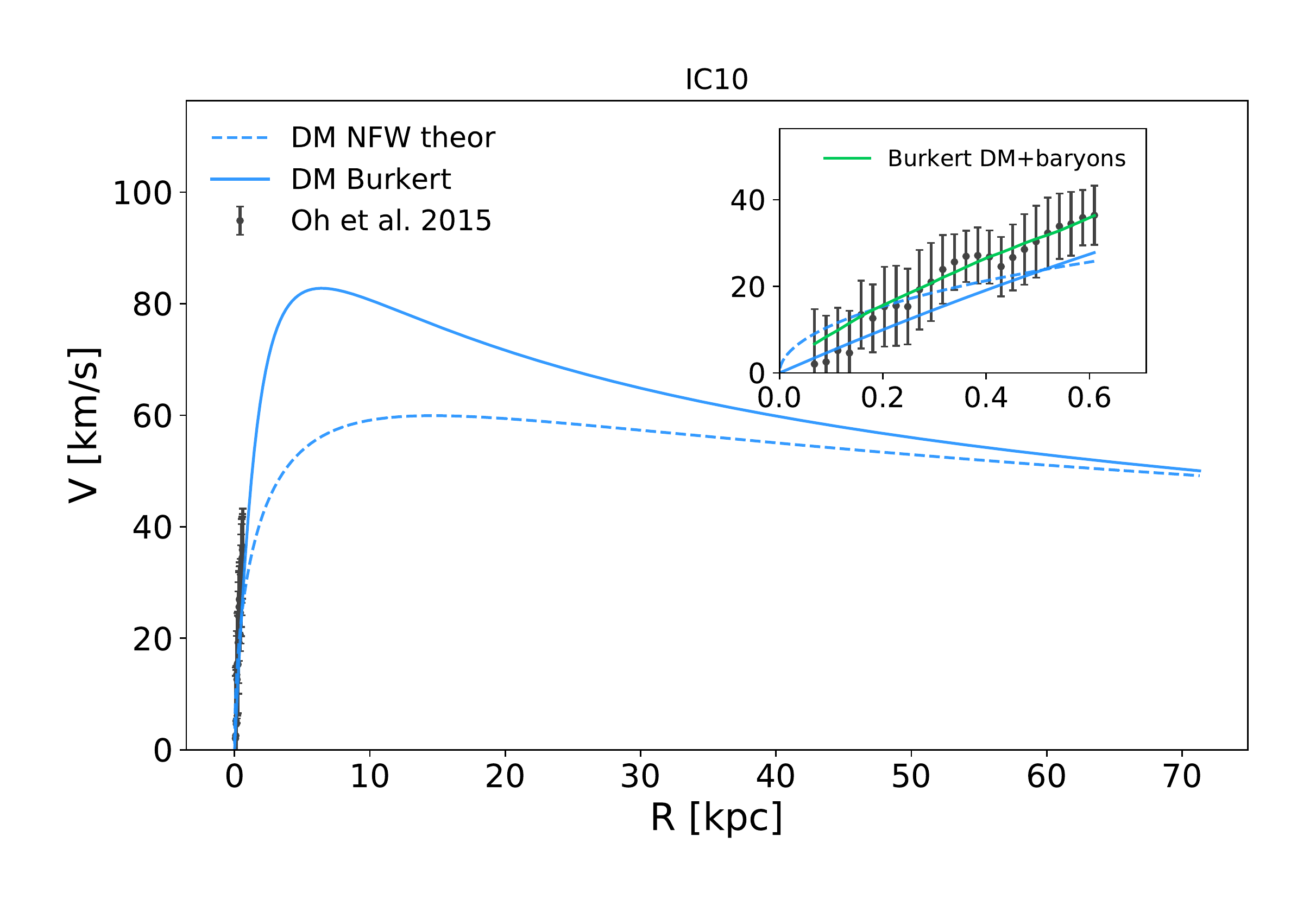}
    \includegraphics[angle=0,width=0.4\linewidth]{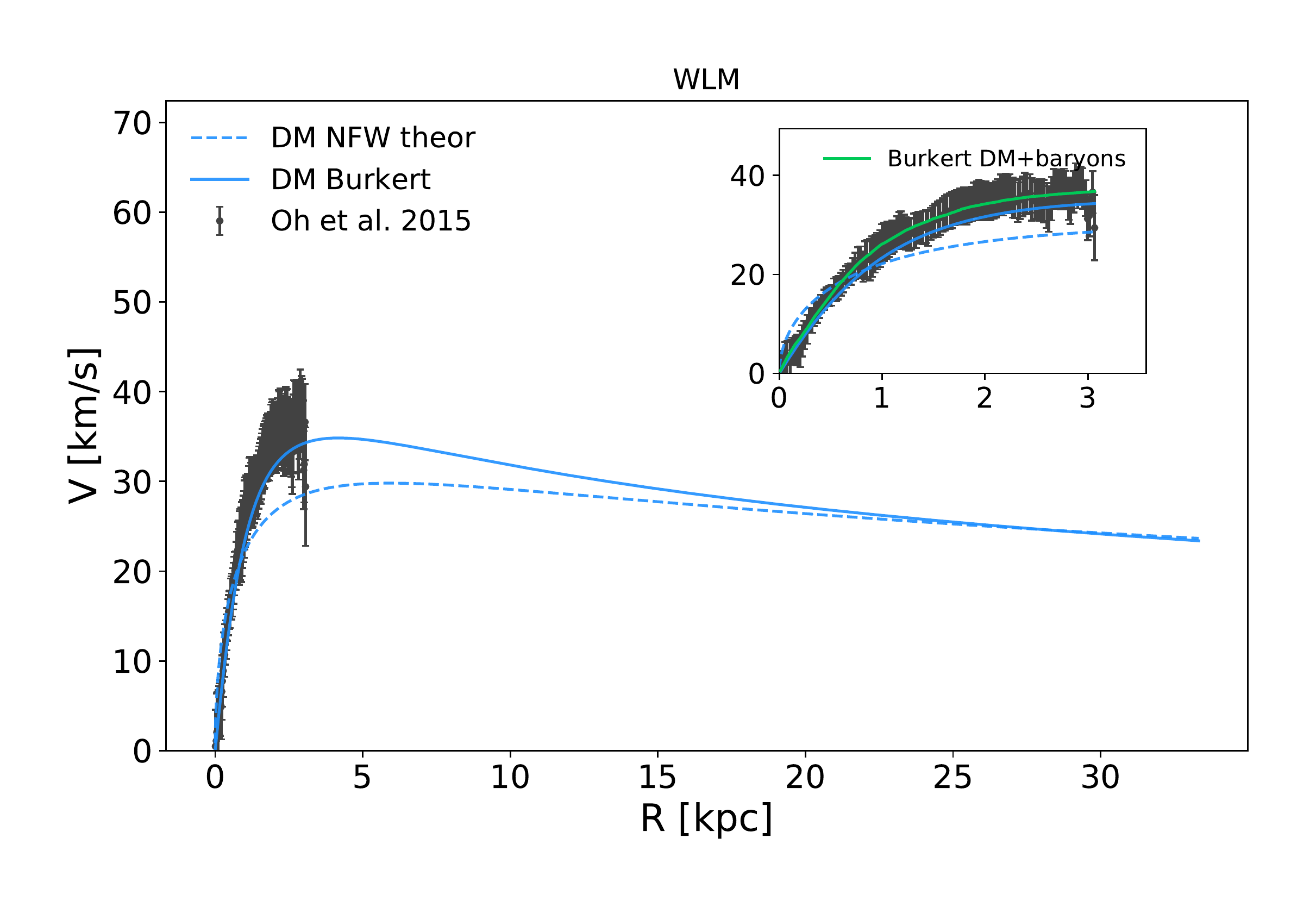}
    \includegraphics[angle=0,width=0.4\linewidth]{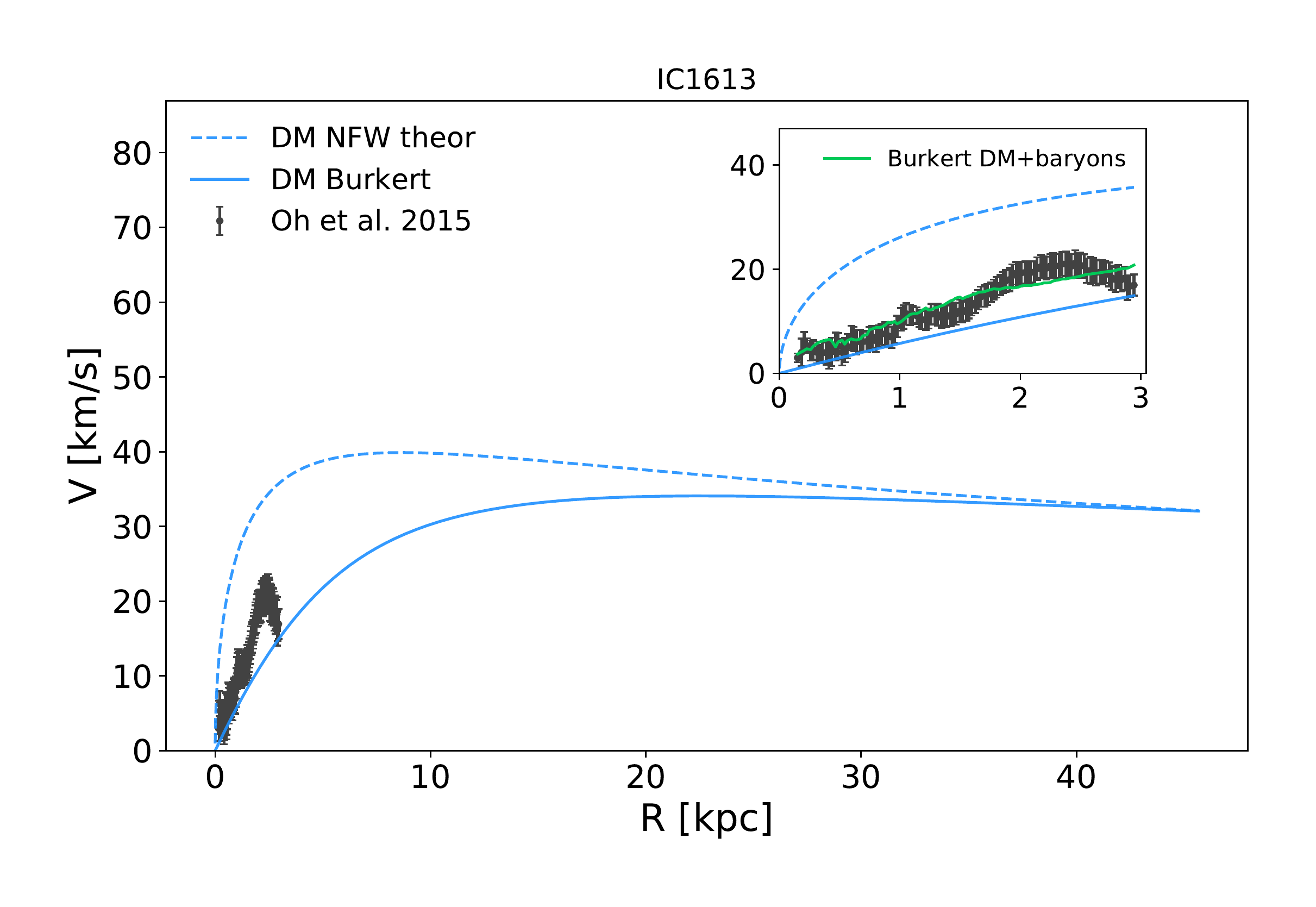}
    \label{RCs}
\end{figure*}


\begin{figure*}
    \centering
    \includegraphics[angle=0,width=0.4\linewidth]{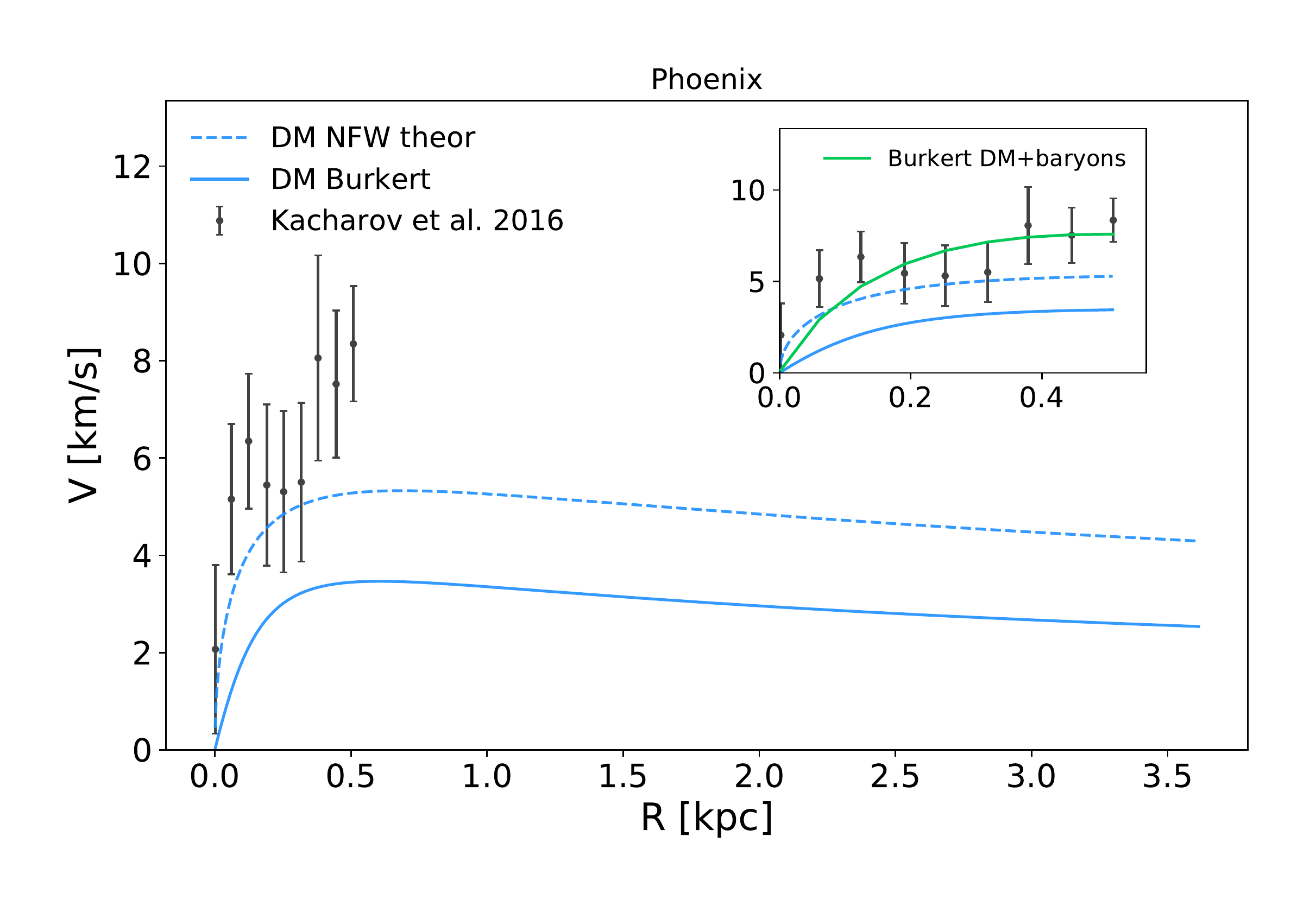}
    \includegraphics[angle=0,width=0.4\linewidth]{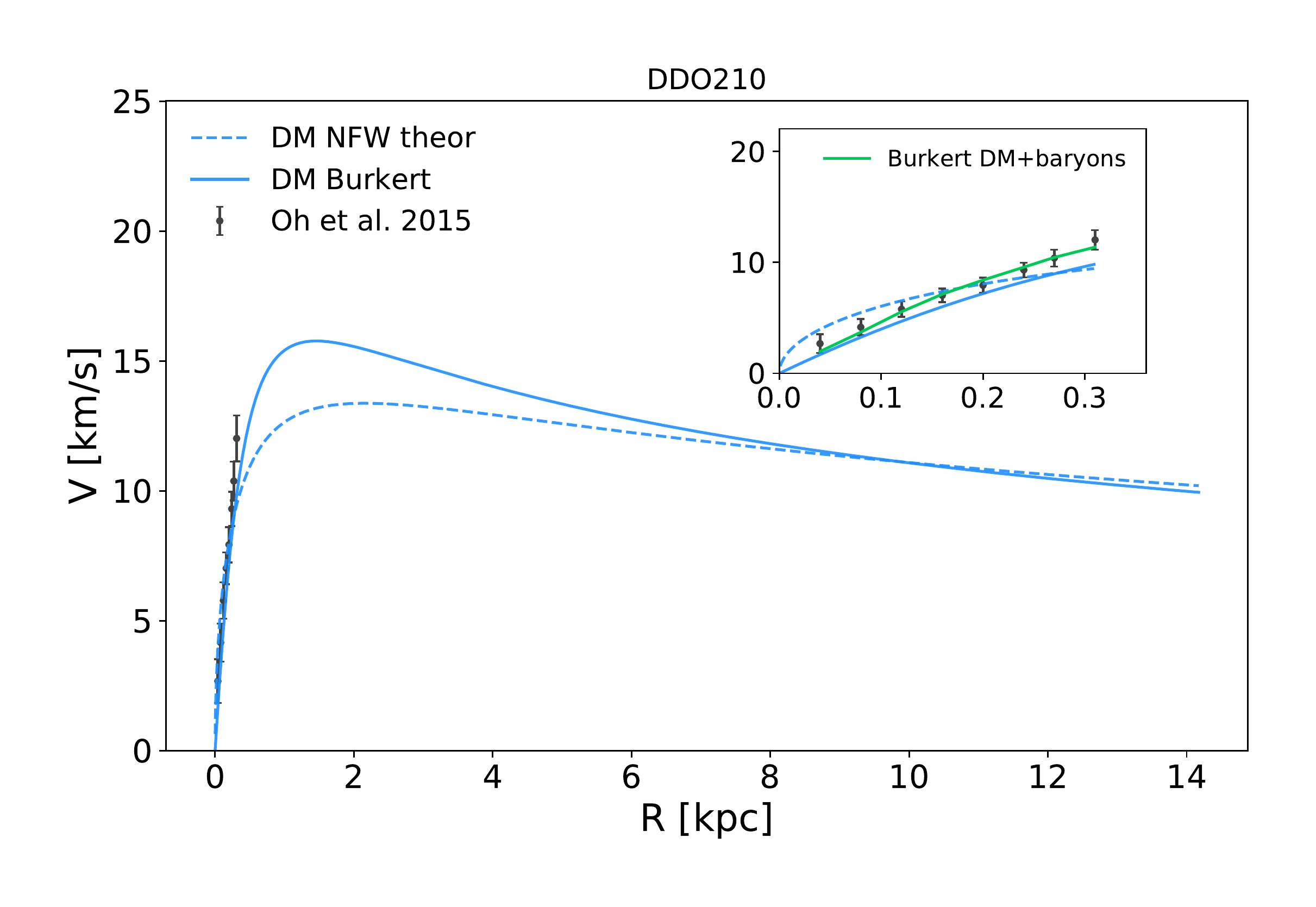}
    \includegraphics[angle=0,width=0.4\linewidth]{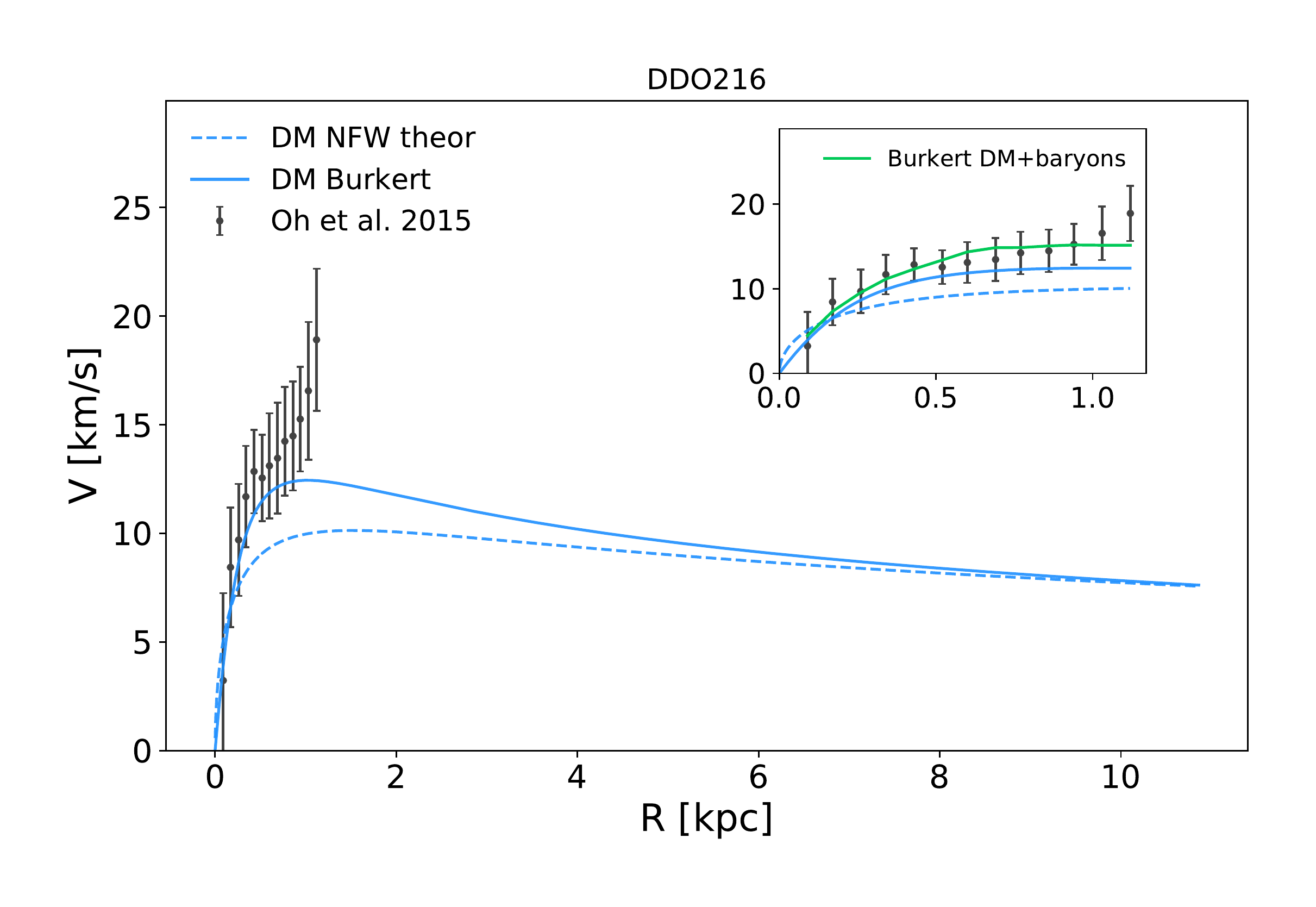}
   \caption{\centering \footnotesize{Rotation curve data plotted along with the fit results taking only the DM component. All the observed data are based on the HI measurements with the exception of Phoenix galaxy for which optical observations are used. Blue solid line represents the Burkert profile while blue dashed the NFW predicted profile (see Section~\ref{Irr} for further details). For each galaxy, the main plots show the RC till $R_{200}$, indeed an indirect representation of consistency of the $M_{200}^\text{Burk}= M_{200}^\text{NFW}$ approximation; in the zoom-in panel we show the results of the total fit (DM+baryons) for the Burkert profile, indeed the discrepancy with the DM-only profile at small radius.
   } }
   \label{RCs}
\end{figure*}



\section{Validation of the pipeline via simulations}
\label{app:simulations}
\begin{figure}[ht!]
    \centering
    \includegraphics[angle=0,width=1\linewidth]{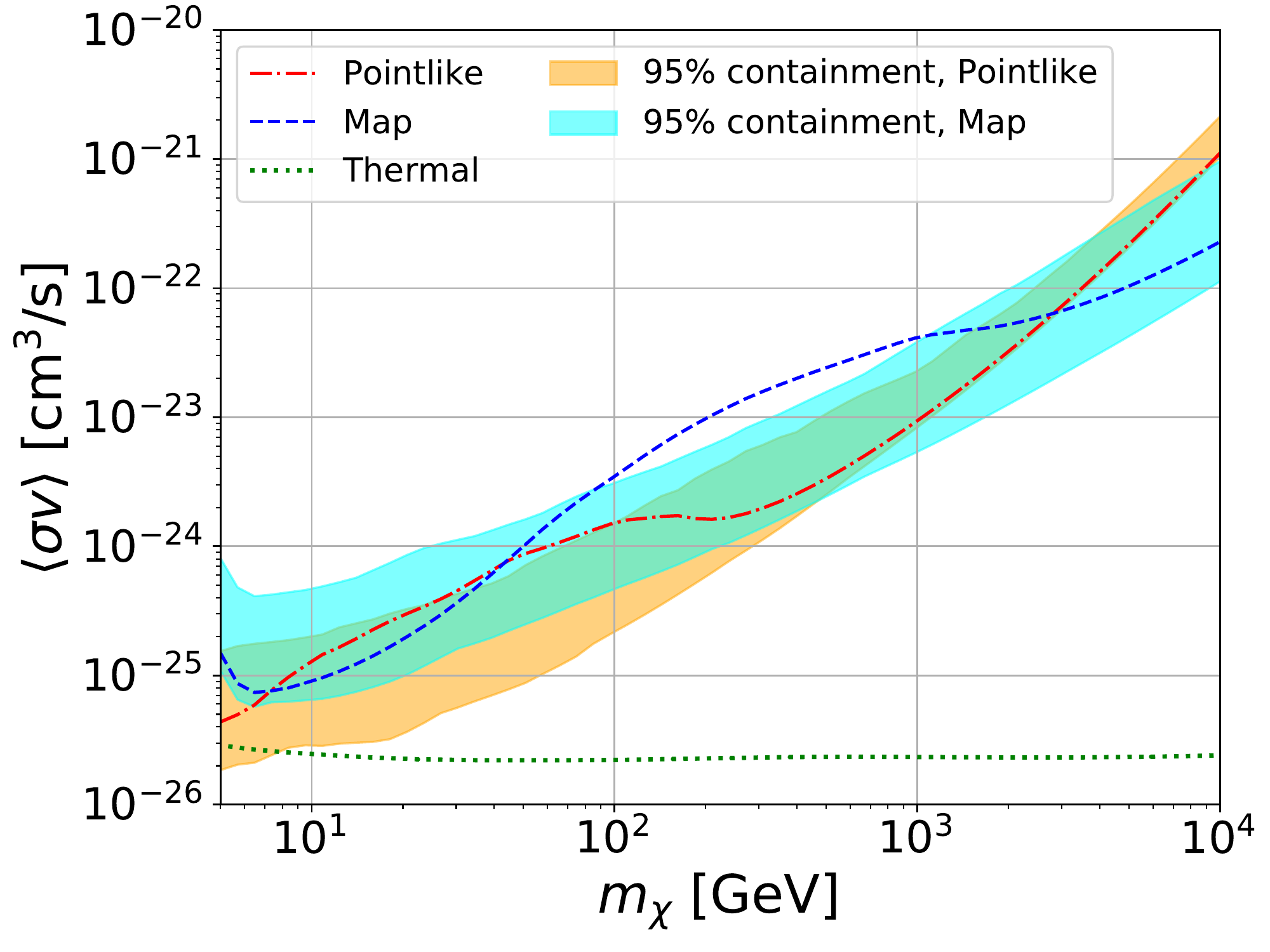}
    \caption{\centering \footnotesize{Control simulations for the $b\bar{b}$ annihilation channel in the MED model. 100 simulations are run to compute the expected DM limits, shown here as the 95\% containment bands for both the null simulation using the spatial template (cyan band) and the point-like analysis in random sky positions (orange band). The actual data constraints are given by the dot-dashed red and dotted blue lines, that were obtained, respectively, for the point-like and extended (spatial template) cases.}}
    \label{fig:null_sim}
\end{figure}

To validate the pipeline, we also run controlled simulations as a check. These are performed assuming i) no DM content in the targets, yet modeled with their spatial templates (null simulations) and ii) random sky pointings (blank fields) assuming point-like sources (i.e., without the spatial DM template). Both scenarios are repeated 100 times, in the first case using 100 pure null simulations and in the second with 100 random sky pointings. The goal of these checks is to compute the expected DM limits in the absence of a DM signal, and to check the robustness of the targets. Both simulations are compatible with our results in Section \ref{sec:DMlimits}, and are shown in Figure \ref{fig:null_sim} for the MED case as an example.There seems to be a potentially significant mismatch at around few hundred GeV between our results with the extended template and the corresponding 95\% c.l. band for the case of using an extended template as well. Yet, we recall that this specific band was computed from null simulations, i.e., no actual data, which is possibly the cause of the observed difference.

In the figure, we see that the results obtained with actual data are contained within the null simulation 95\% c.l. uncertainty band, except for the 100--1000 GeV region, where the actual limits are degraded due to the found excesses reported and discussed in Section \ref{gammarayDMsection}. On the other hand, the random pointing results are compatible with the point-like simulations at 95\% c.l., which validates our analysis pipeline.

\section{Flux upper limits}
\label{app:flux_upper_limits}
\begin{figure*}
    \centering
    \includegraphics[width=0.40\linewidth]{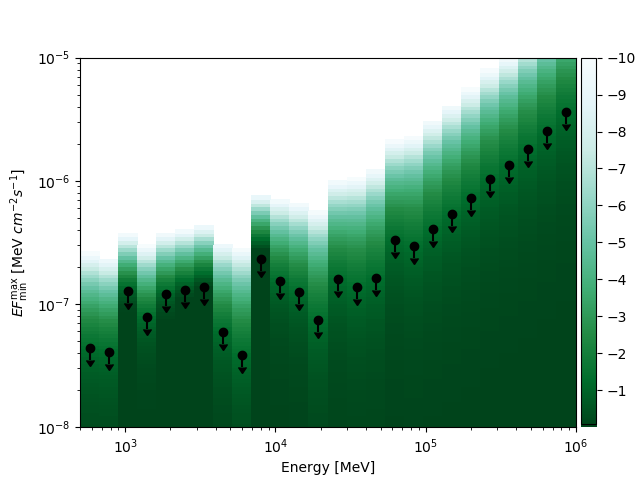}
    \includegraphics[width=0.40\linewidth]{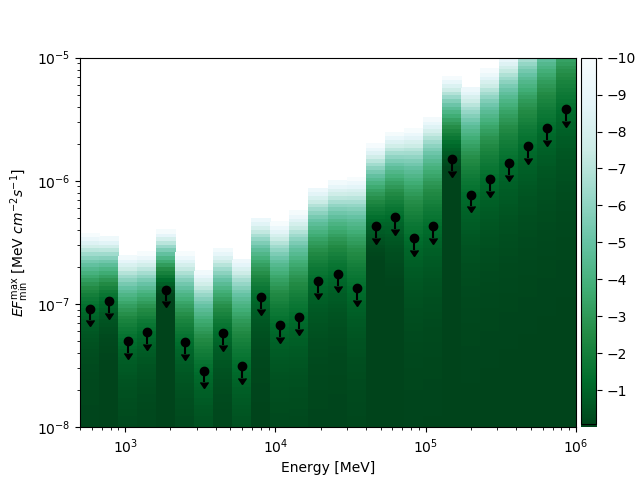}
    \includegraphics[width=0.40\linewidth]{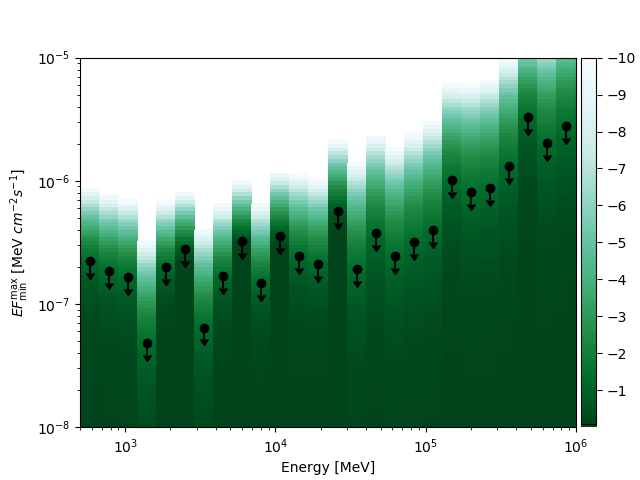}
    \includegraphics[width=0.40\linewidth]{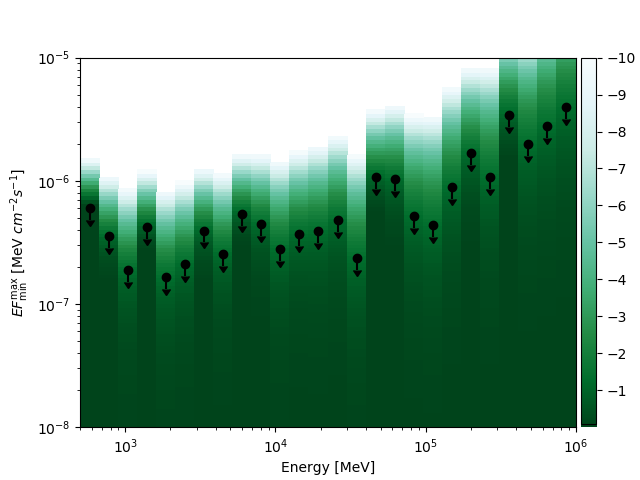}
    \includegraphics[width=0.40\linewidth]{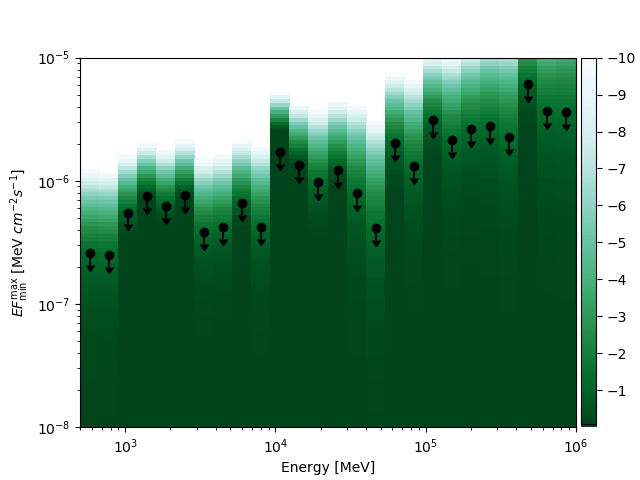}
    \includegraphics[width=0.40\linewidth]{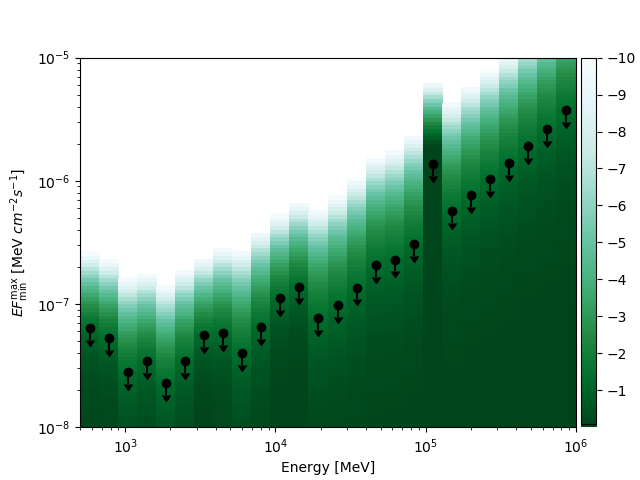}
    \includegraphics[width=0.40\linewidth]{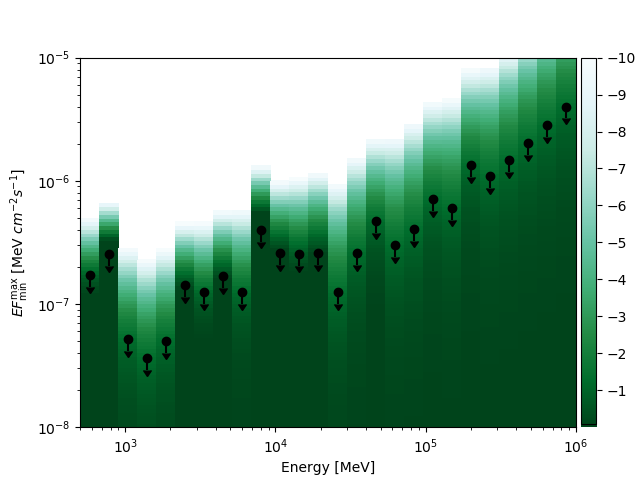}
    \caption{\centering \footnotesize{95\% C.L. upper limits to the flux and likelihood values found for the dIrrs in the gamma-ray analysis performed in section \ref{gammarayDMsection} with the \textit{Fermi}-LAT. Color code traces the change in log-likelihood. The arrows indicate upper limits, as no signal is detected in any of the bins. Panels show, from left to right and top to bottom, DDO210, DD0216, IC10, IC1613, NGC6822, Phoenix, and WLM.}}
    \label{fig:flux_upper_limits}
\end{figure*}

In this Appendix, we show the 95\% C.L. flux upper limits as measured by the LAT instrument, with the analysis setup explained in Section \ref{gammarayDMsection}. The limits are shown in Figure \ref{fig:flux_upper_limits}.

\section{Individual DM limits for $\tau^+\tau^-$ and $W^+W^-$ channels}
\label{app:individual_limits_tt_ww}
\begin{figure*}
\centering
\includegraphics[angle=0,width=0.40\linewidth]{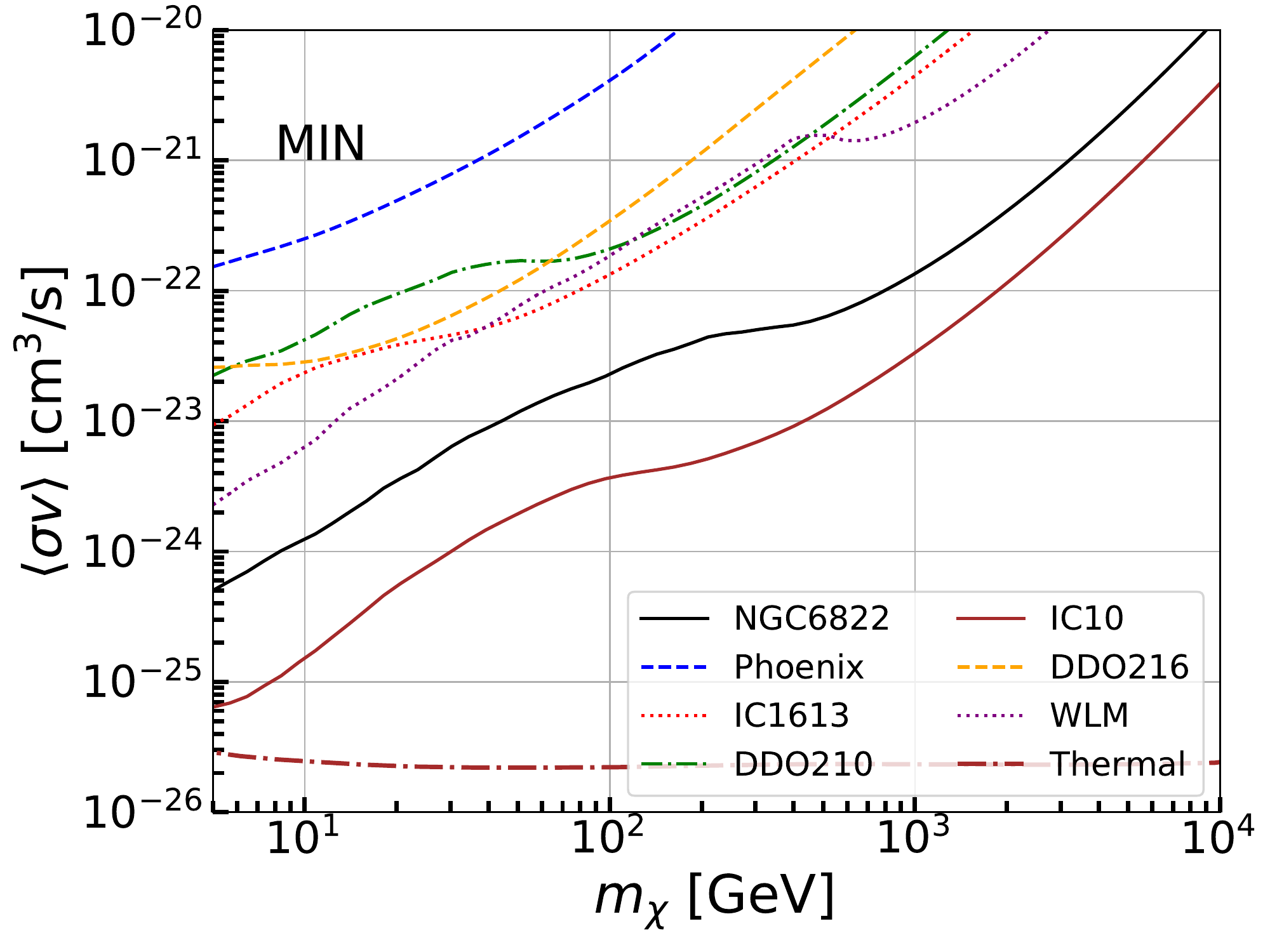}
\includegraphics[angle=0,width=0.40\linewidth]{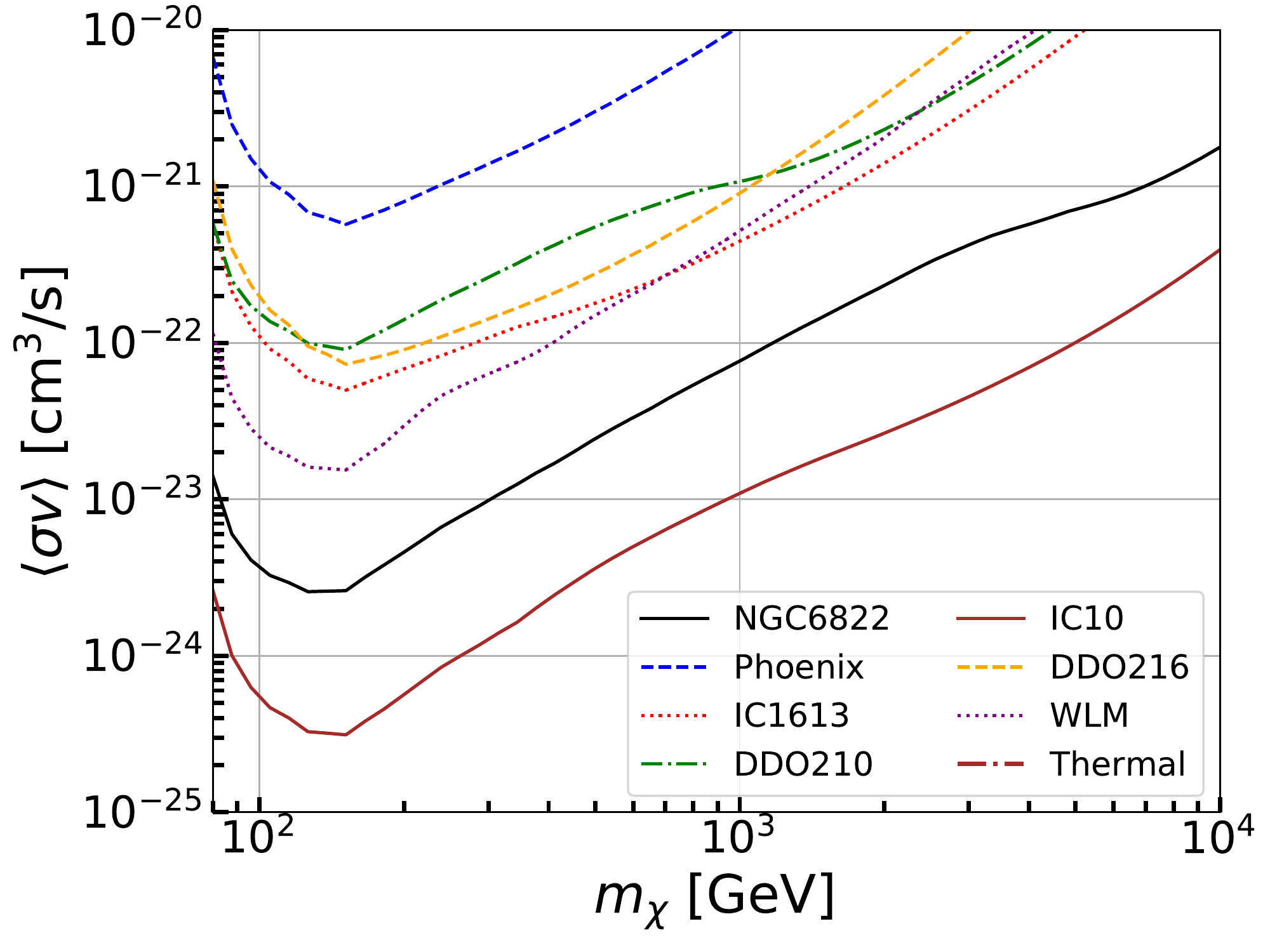}
\includegraphics[angle=0,width=0.40\linewidth]{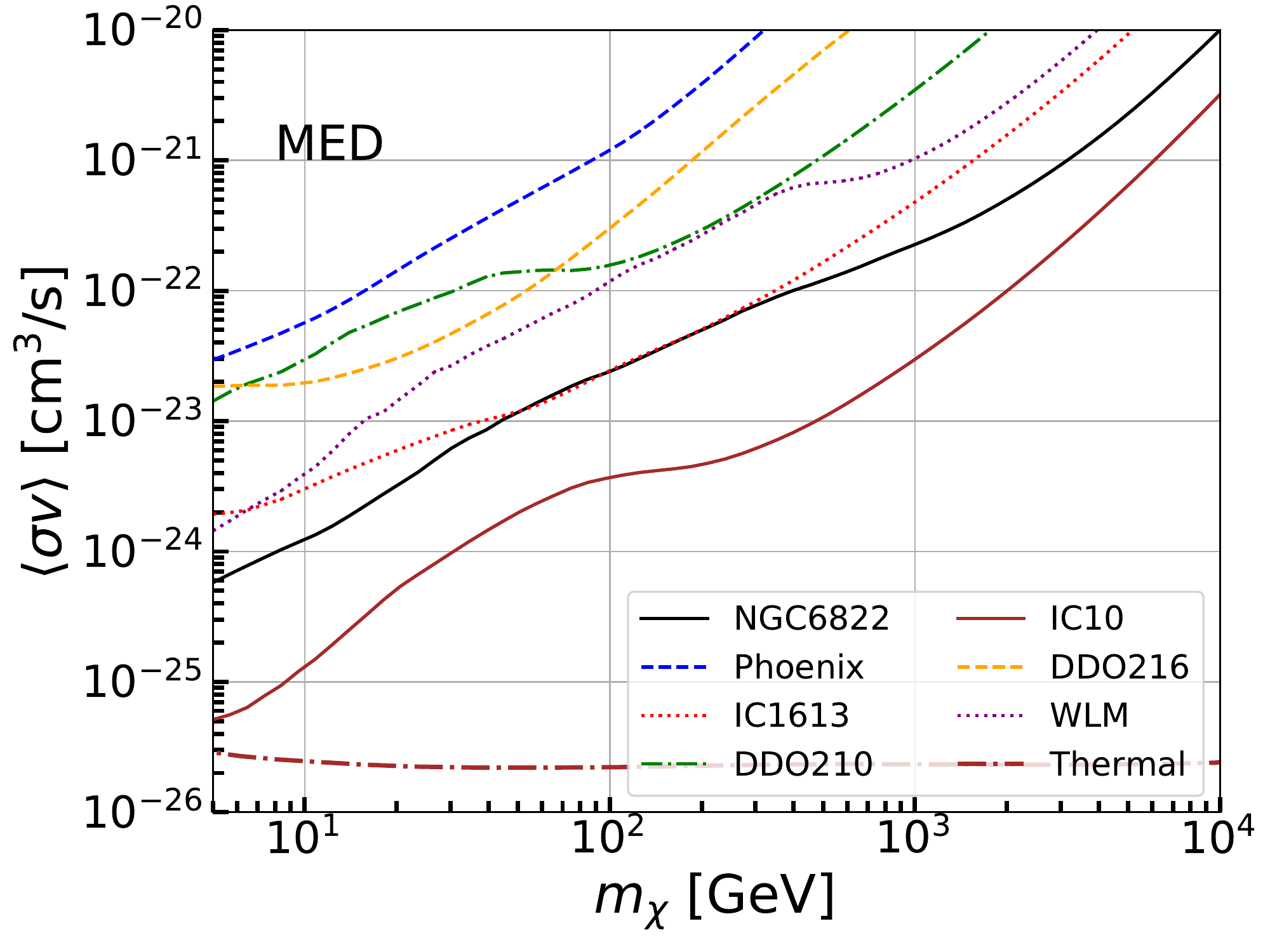}
\includegraphics[angle=0,width=0.40\linewidth]{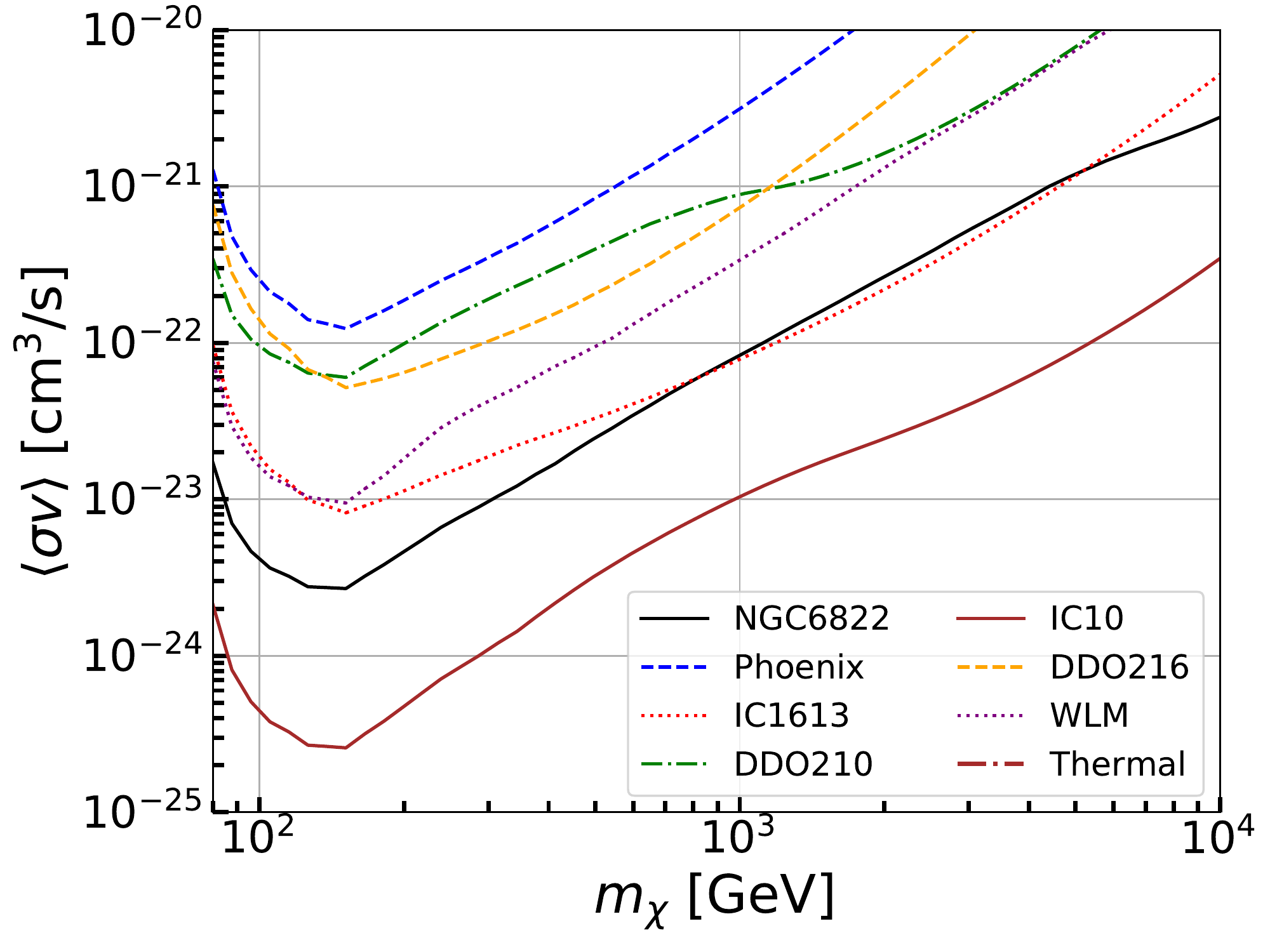}
\includegraphics[angle=0,width=0.40\linewidth]{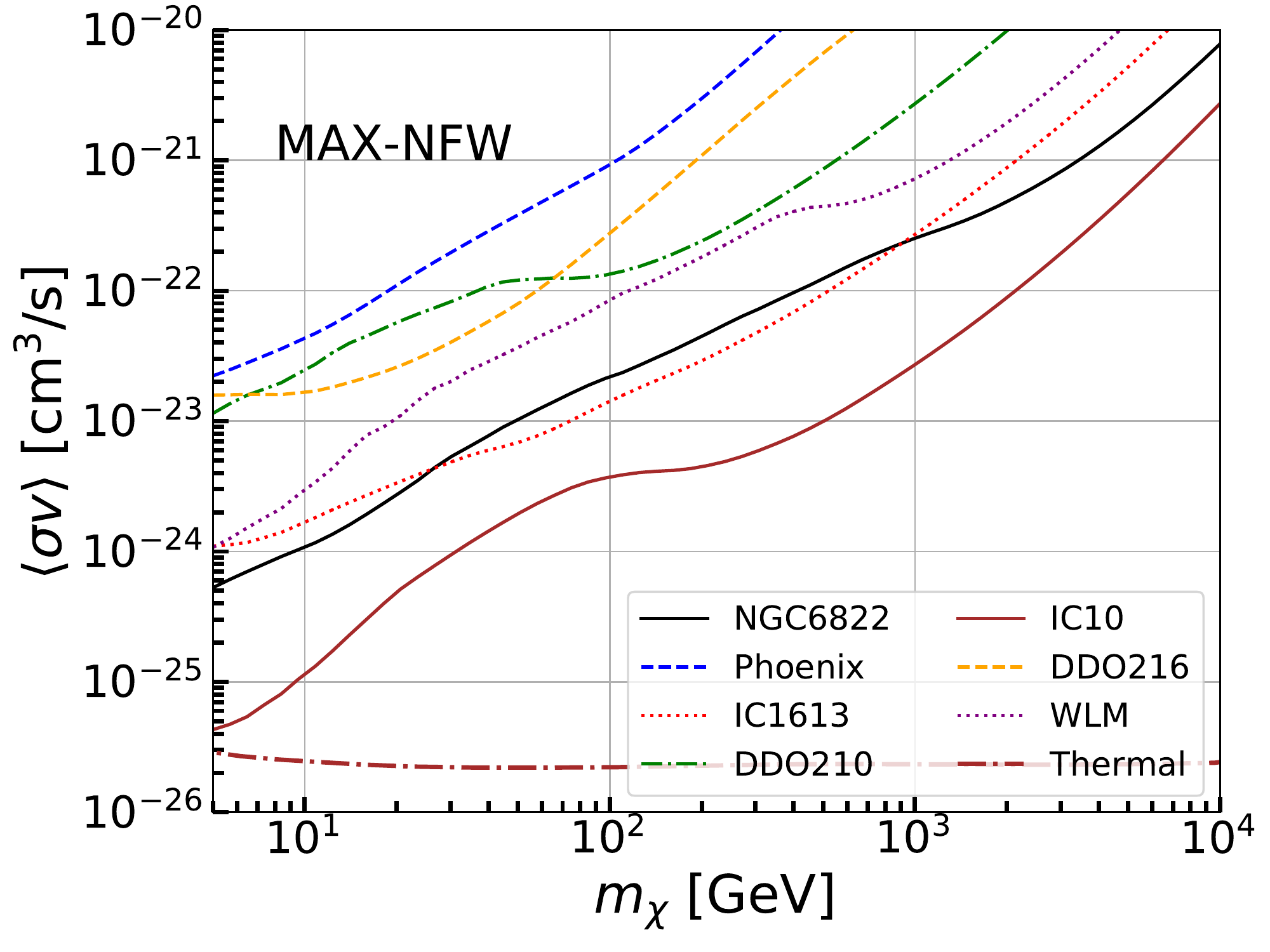}
\includegraphics[angle=0,width=0.40\linewidth]{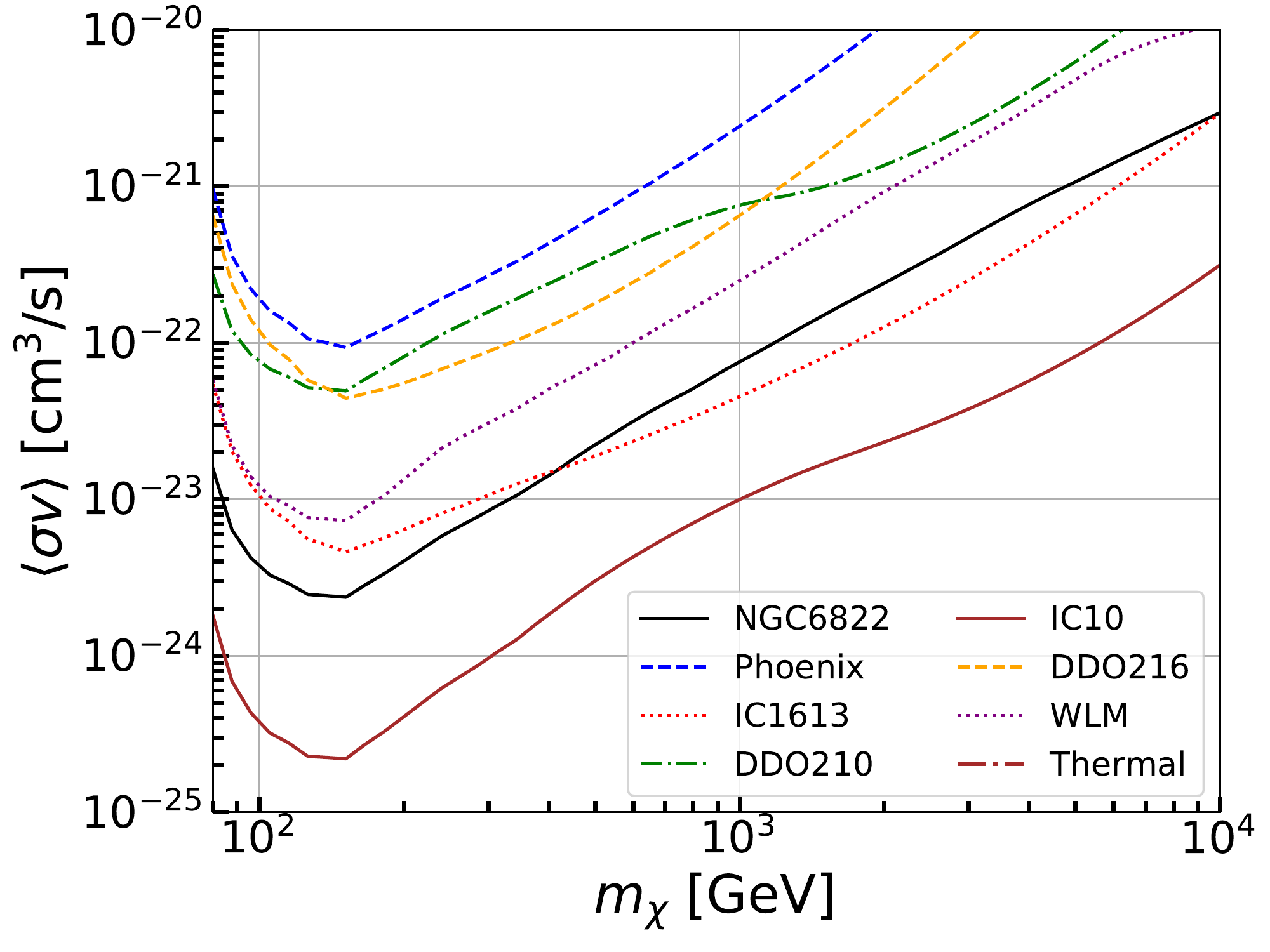}
\includegraphics[angle=0,width=0.40\linewidth]{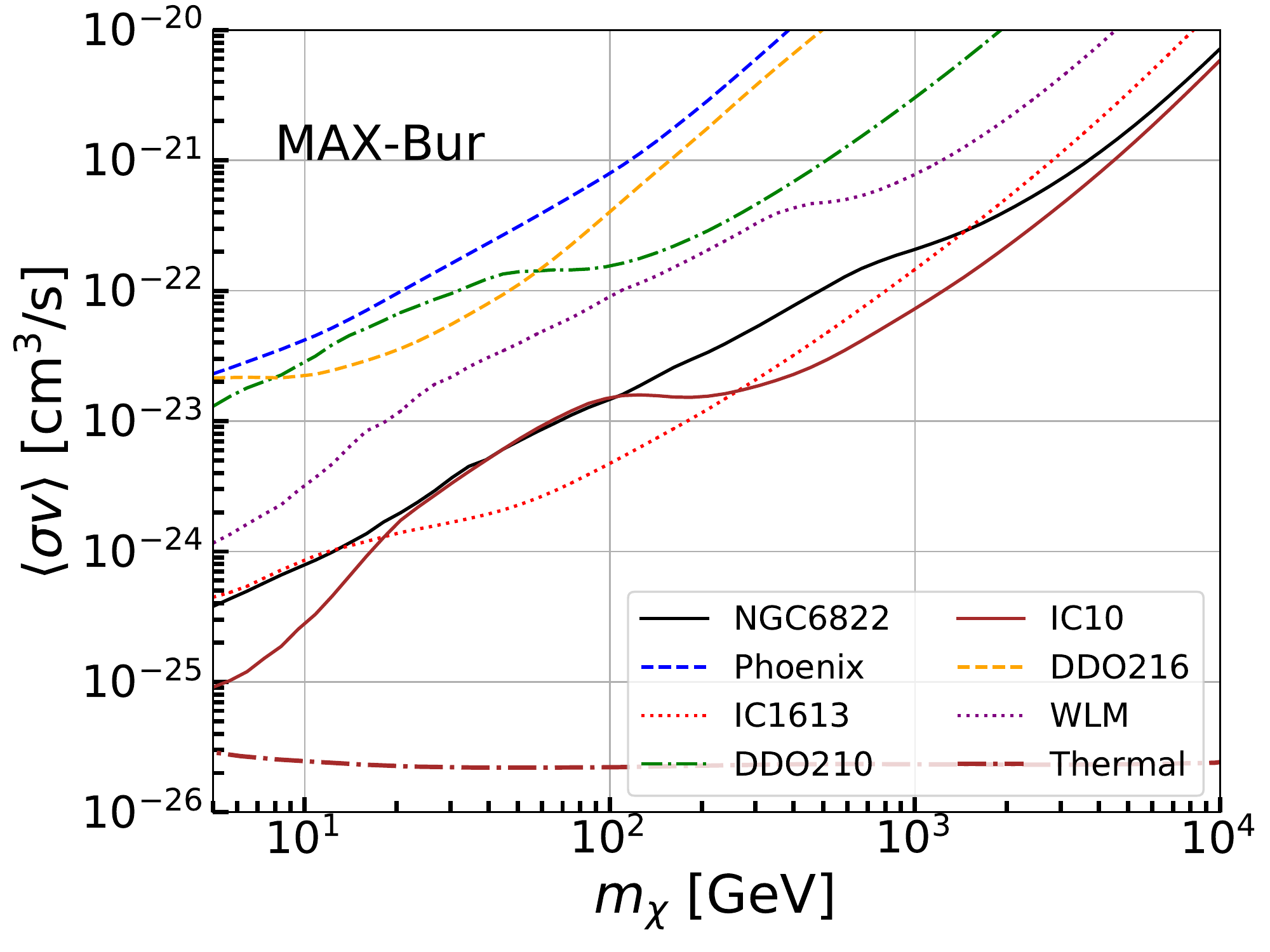}
\includegraphics[angle=0,width=0.40\linewidth]{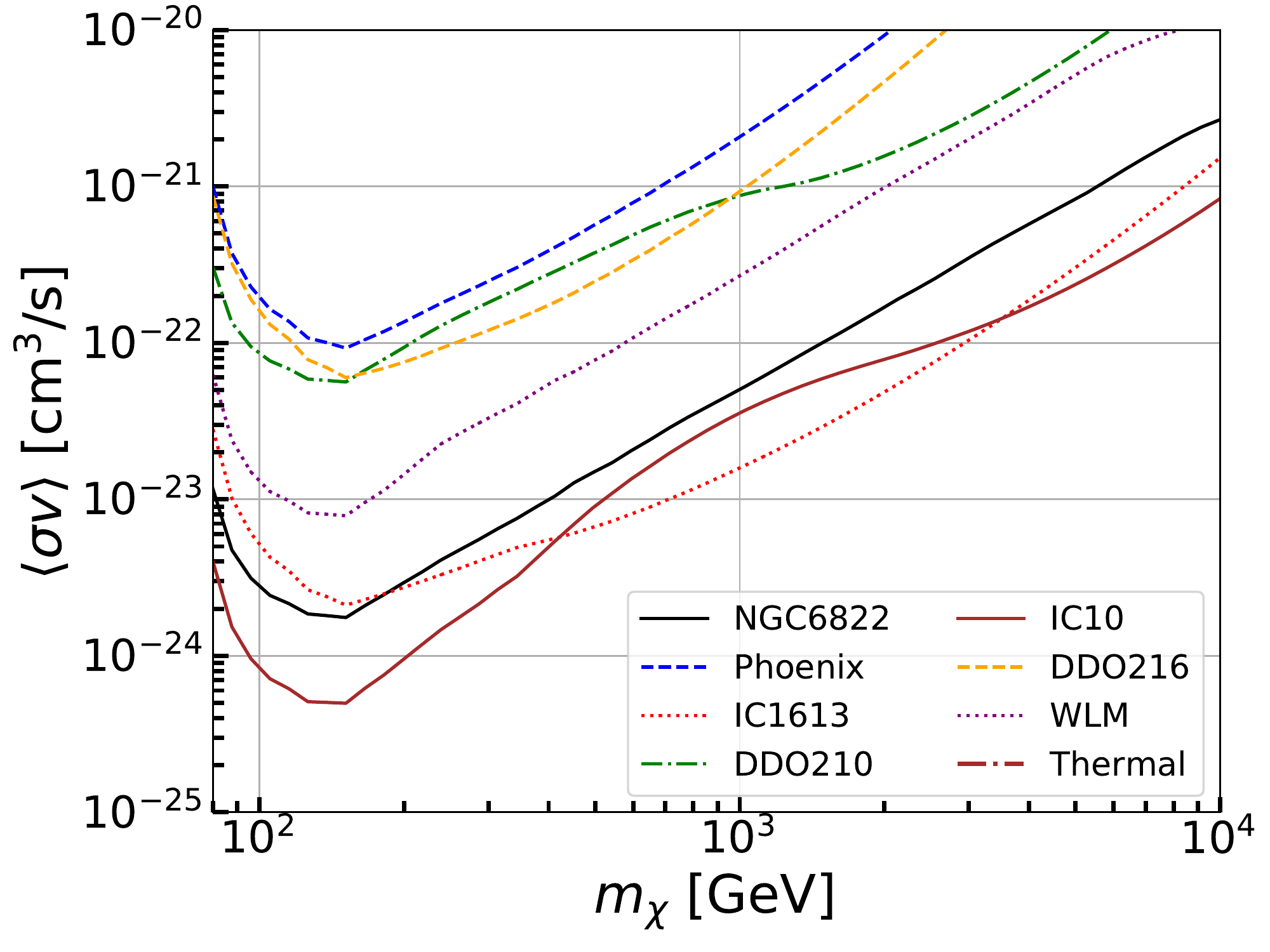}
\caption{\centering \footnotesize{\textbf{Left column:} Upper limits for $\langle\sigma v\rangle$ to each individual source, for the $\tau^+\tau^-$ annihilation channel. The different models considered are, from top to bottom, MIN, MED, MAX-Bur and MAX-NFW. \textbf{Right column:} Same as 
left column but for the $W^+W^-$ annihilation channel.
}}
\label{fig:tautau_WW_individual}
\end{figure*}

In Figure \ref{fig:bb_individual} of the main text, we showed  individual DM limits for each dIrr in our sample and DM model for the case of annihilations to $b\bar{b}$. In this Appendix, we also show DM limits for the other two considered channels, $\tau^+\tau^-$ and $W^+W^-$, in Figure \ref{fig:tautau_WW_individual}.

\newpage

%






\end{document}

%% file: PRD_Fermi_v1.bbl
%